\documentclass[11pt,tightenlines,superscriptaddress,nofootinbib]{revtex4}

\usepackage[page,title]{appendix}
\usepackage{url}
\usepackage{placeins}
\usepackage{amsfonts}
\usepackage[dvipsnames]{xcolor}
\usepackage{color}
\usepackage{xspace}
\usepackage{tikz}
\usepackage{feynmf}
\usepackage[utf8]{inputenc}											
\usepackage{subfigure}

\usetikzlibrary{arrows}
\usetikzlibrary{shapes}
\usetikzlibrary{trees}
\usetikzlibrary{matrix}
\usetikzlibrary{arrows} 			
\renewcommand{\draft}[1]{{\color{blue}#1}}

\usepackage{graphicx,psfrag,bm}
\usepackage{mathrsfs,amsfonts}
\usepackage{epsfig,cancel}
\usepackage{epstopdf}
\usepackage{mciteplus}
\usepackage{latexsym}
\usepackage{amsthm}
\usepackage{amsmath}
\usepackage{amssymb}
\usepackage{hepunits}
\usepackage{hyperref}
\usepackage{bbm}
\usepackage{xfrac}
\usepackage{colordvi}
\usepackage{comment}
\usepackage{dcolumn}
\usepackage{times,wrapfig}
\usepackage{slashed}
\usepackage{booktabs}
\usepackage{relsize}
\newcommand{\pb}{\,\,\ensuremath{\mathrm{pb}}}
\newcommand{\ifb}{\,\,\ensuremath{\text{fb}^{-1}}}
\newcommand{\iab}{\,\,\ensuremath{\text{ab}^{-1}}}
\newcommand{\Tev}{\,\,\ensuremath{\mathrm{TeV}}}
\newcommand{\Gev}{\,\,\ensuremath{\mathrm{GeV}}}
\newcommand{\Mev}{\,\,\ensuremath{\mathrm{MeV}}}

\newcommand{\mll}{\ensuremath{m_{\ell\ell}}}
\newcommand{\zdark}{\ensuremath{Z_D}}
\newcommand{\ptmiss}{\ensuremath{p_{T\text{,miss}}}}
\newcommand{\ndark}{\ensuremath{N_D}}
\newcommand{\UmFour}{\ensuremath{U_{\mu4}}}
\newcommand{\pt}{\ensuremath{p_{\text{T}}}}
\def\babar{\mbox{\slshape B\kern-0.1em{\smaller A}\kern-0.1emB\kern-0.1em{\smaller A\kern-0.2em R}}}

\newcommand{\ednote}[1]{} 
\newcommand{\people}[1]{} 
\newcommand{\morepeople}[1]{} 

\colorlet{RED}{red}
\colorlet{BLUE}{blue}
\colorlet{ORANGE}{orange}

\begin{document}

\title{Shedding Light on the MiniBoone Excess with Searches at the LHC}

\date{\today}
\hspace*{0pt}\hfill 
{\small  FERMILAB-PUB-23-516-CMS-CSAID-PPD-T}

\bigskip

\author{Christian~Herwig}
\affiliation{Fermi National Accelerator Laboratory, Batavia, IL 60510, USA}

\author{Joshua~Isaacson}
\affiliation{Fermi National Accelerator Laboratory, Batavia, IL 60510, USA}

\author{Bo~Jayatilaka}
\affiliation{Fermi National Accelerator Laboratory, Batavia, IL 60510, USA}

\author{Pedro~A.~N.~Machado}
\affiliation{Fermi National Accelerator Laboratory, Batavia, IL 60510, USA}

\author{Allie~Reinsvold~Hall}
\affiliation{Fermi National Accelerator Laboratory, Batavia, IL 60510, USA}
\affiliation{United States Naval Academy, Annapolis, MD 21402, USA}

\author{Murtaza~Safdari}
\affiliation{Fermi National Accelerator Laboratory, Batavia, IL 60510, USA}

\begin{abstract}

The origin of the excess of low-energy events observed by the MiniBooNE experiment remains a mystery, despite exhaustive investigations of backgrounds and a series of null measurements from complementary experiments.
One intriguing explanation is the production of beyond-the-Standard-Model particles that could mimic the experimental signature of additional $\nu_e$ appearance seen in MiniBooNE.
In one proposed mechanism, muon neutrinos up-scatter to produce a new ``dark neutrino'' state that decays by emitting highly-collimated electron-positron pairs.
We propose high-energy neutrinos produced from $W$ boson decays at the Large Hadron Collider as an ideal laboratory to study such models.
Simple searches for a low-mass, boosted di-lepton resonance produced in association with a high-$p_\text{T}$ muon from the $W$ decay with Run~2 data would already provide unique sensitivity to a range of dark neutrino scenarios, with prompt and displaced searches providing complementarity.
Looking farther ahead, we show how the unprecedented sample of $W$ boson decays anticipated at the HL-LHC, together with improved lepton acceptance would explore much of the parameter space most compatible with the MiniBooNE excess.



\end{abstract}

\maketitle
\tableofcontents

\section{Introduction}

For several decades, the MiniBooNE experiment has presented an unresolved anomaly in its data, characterized by the observation of an excess of electron-like events over the predicted background in a muon-neutrino-dominated beam~\cite{MiniBooNE:2007uho, MiniBooNE:2020pnu}. 
Motivated by other short baseline neutrino anomalies, such as the LSND result~\cite{LSND:2001aii}, or the gallium anomaly~\cite{Barinov:2022wfh}, several proposed explanations for the MiniBooNE low-energy excess rely on oscillations or neutrino flavor transitions in general.
Nevertheless, these explanations typically face strong constraints from other neutrino oscillation experiments~\cite{Collin:2016rao, Gariazzo:2017fdh, Dentler:2018sju}, or from cosmological observations~\cite{Planck:2018vyg}.

Due to these difficulties, other directions on how to address the MiniBooNE anomaly have been explored~\cite{Gninenko:2009ks, Bertuzzo:2018itn, Bertuzzo:2018ftf, Arguelles:2018mtc, Ballett:2018ynz, Fischer:2019fbw, Datta:2020auq, Abdallah:2020biq, Dutta:2021cip}.
These use the fact that what MiniBooNE actually measures is the Cherenkov light produced by charged particles as they traverse its mineral oil detector. 
Therefore, signals that could produce electron-like Cherenkov rings have the potential to explain the anomaly. 

Despite many attempts to explain the anomaly, its resolution is still an open problem. 
One intriguing model that has recently been proposed as a possible solution is the ``dark neutrino" model~\cite{Bertuzzo:2018itn, Bertuzzo:2018ftf, Ballett:2018ynz}. 
In this scenario, the breaking of a new gauge symmetry acting only in Standard Model (SM) singlets, i.e., in the dark sector, leads to the smallness of neutrino masses. 
This mechanism naturally leads to mixing between standard model neutrinos and dark neutrinos, which generates a small interaction among these particles and the new gauge boson.    
This new interaction leads to novel phenomenology at MiniBooNE: muon neutrinos can up-scatter to dark neutrinos via this new interaction, followed by the decay of the dark neutrinos into a light neutrino and an electron-positron pair. 
If the $e^+e^-$ pair is sufficiently collimated, or if there is a large energy asymmetry in these final state particles, the MiniBooNE detector would classify it as an electron-like signal, thus providing a possible explanation for the anomaly.

There have been several studies on the viability of these scenarios to address the MiniBooNE anomaly, see e.g. Refs.~\cite{Bertuzzo:2018itn, Bertuzzo:2018ftf, Ballett:2018ynz, Arguelles:2018mtc, Datta:2022zng,
 Datta:2023iln}~\footnote{Flavor constraints discussed in Ref.~\cite{Datta:2023iln} rely on large $Z$-$Z_D$ mass mixing which in ultraviolet complete realizations of the dark neutrino model tends to be very small~\cite{Bertuzzo:2018ftf}.}.
Notably, the angular distribution of the Cherenkov ring in electron-like events detected in MiniBooNE seems to point to an excess that is not concentrated only in the forward direction.
At the same time, theoretical challenges in modeling the angular distribution of the background temper the ability to draw strong conclusions from these data.
Currently, discrepancies between data and the calculated leptonic angular distribution have been observed in multiple experiments, even after tuning predictions to data. 
For example, recent MicroBooNE~\cite{MicroBooNE:2021nxr} and ArgoNeuT analyses~\cite{Duffy:2021hie} each predict yields that are consistently above the data, in excess of the $\sim20\%$ uncertainty band.
Another MicroBooNE analysis~\cite{MicroBooNE:2023cmw} reveals that normalizing the theoretical prediction to match the number of events observed would lead to a large over-prediction of forward events even for the tuned generator.
Even in MiniBooNE itself~\cite{Katori:2008zz}, before tuning the estimate from simulation, the prediction misses the data significantly in the forward region.

Additionally, the dark neutrino scenario has not yet been interfaced with realistic models of the target nuclei. 
In a slightly different context, it has been shown~\cite{Kopp:2024yvh} that this interface can significantly change the beyond-the-SM cross section. 
Similar effects may in turn change the dark neutrino region of parameter space preferred by MiniBooNE once fully accounted for.
There has also been a recasting of old CHARM-II neutrino-electron scattering data~\cite{Arguelles:2018mtc} that may disfavor part of the MiniBooNE preferred region, but these types of analyses suffer from the same unknowns discussed above.

The MicroBooNE experiment has released their first set of results attempting to further investigate the MiniBooNE anomaly~\cite{MicroBooNE:2021tya,MicroBooNE:2021nxr,MicroBooNE:2021pvo,TheMicroBooNECollaboration:2021cjf}.
In these analyses, MicroBooNE requires the final state to have exactly one electron/positron.
For the dark neutrino model, while the $e^+e^-$ pair would be reconstructed as an electron-like signal at MiniBooNE, the signature at MicroBooNE could be quite different, since liquid argon detectors have much better particle reconstruction capabilities.
Therefore, while the results from MicroBooNE are a great first step in addressing the MiniBooNE anomaly, they do not rule out the sterile neutrino interpretation of the MiniBooNE excess~\cite{Arguelles:2021meu, MicroBooNE:2022sdp}, neither models with multiple lepton final states, nor other possible explanations of the excess~\cite{Arguelles:2019xgp}.
Given this current situation, it would be invaluable to have an orthogonal test of the dark neutrino scenario, independent of neutrino-nucleus interaction modeling uncertainties, such as the LHC search we propose here.

From a theoretical view, the dark neutrino model is an interesting model of low-scale neutrino mass that could lead to observable phenomenology in laboratory experiments~\cite{Bertuzzo:2018ftf, Ballett:2019pyw}. 
Given the lack of experimental guidance on the scale of the physics that is responsible for neutrino mass generation, it is crucial to develop searches that could be sensitive to both high-scale models, such as GUT-embedded seesaw scenarios, or low-scale models such as the dark neutrino scenario. 
Although the dark neutrino phenomenology has been studied in neutrino and meson decay experiments, possible signatures at the LHC have not received much attention.

Dark neutrinos could be copiously produced by the decay of on-shell $W$ bosons.
With a $W$ production cross-section of $\sigma(pp\to W)=190\text{ nb}$ at $\sqrt{s}=13\text{ TeV}$ center of mass energy~\cite{ATLAS:2016fij,Anastasiou:2003ds}, the LHC has produced roughly 30 billion $W$ bosons during Run~2 data-taking at each of the ATLAS and CMS experiments.
The ultimate High-Luminosity LHC (HL-LHC) should enlarge these samples by an additional factor of 20-30, collecting an integrated luminosity up to $4\text{ ab}^{-1}$ and profiting from the slightly larger cross section at $\sqrt{s}=14\text{ TeV}$.
Therefore, even a small active-dark neutrino mixing could lead to a significant production of these new particles at the LHC. 
The dark neutrino will then decay into an $e^+e^-$ pair, which would be the smoking gun signature of this model. 
While probing the very low-mass end of this scenario may be challenging at the LHC due to a large off-shell photon-to-$e^+e^-$ background, the LHC provides a complementary probe of such mass models, being sensitive to the hundred-MeV to multi-GeV scale.
In this mass range, other decay modes could also open up, such as dimuon or hadronic final states.

In this work, we explore the connection between LHC physics and neutrino experiments, highlighting an interesting complementarity between these high and low energy scale probes of new physics. 
We present a detailed study of the dark neutrino model in the context of LHC searches, showing how the HL-LHC can probe models of light new physics. 
We develop a new LHC search that can probe parameter space relevant to the MiniBooNE excess, and set leading constraints for a significant portion.

The paper is structured as follows Sec.~\ref{sec:theory} reviews the dark neutrino model and Sec.~\ref{sec:concept} details the search strategy at the LHC.
The simulation of events along with event reconstruction is discussed in Sec.~\ref{sec:mc}.
An analysis of the possible constraints that the LHC could set on the dark neutrino model is carried out in Sec.~\ref{sec:analysisW}, with results shown in Sec.~\ref{sec:results}.
Finally, Sec.~\ref{sec:outlook} provides some outlook to other possible searches that could be carried out in this low energy region and summarizes the findings of this work.

\section{Dark Neutrino Model}
\label{sec:theory}

The dark neutrino model consists of a realization of the neutrino mass mechanism at low scales~\cite{Bertuzzo:2018ftf}.
In a nutshell, a new $U(1)_D$ gauge symmetry acting only on SM singlets would lead to the conservation of lepton number if left unbroken.
These ``dark neutrinos'' are vector-like under this new symmetry to ensure anomaly cancellation. 
The spontaneous breaking of this symmetry by scalar fields in the dark sector simultaneously breaks lepton number, which leads to nonzero neutrino masses.
The breaking of the dark symmetry mixes active and dark neutrinos, leading to a new interaction of active neutrinos: an up-scattering to dark neutrinos via the exchange of the new gauge boson, $Z_D$, which kinetically mixes with the photon.
This new interaction could potentially explain the MiniBooNE anomaly~\cite{Bertuzzo:2018itn}: the dark neutrinos would decay to light neutrinos and $e^+e^-$ pairs; if the $e^+e^-$ opening angle is small enough, MiniBooNE would classify these events as electron-like events, \textit{i.e.} the signature of the anomaly.

To be more concrete, let us describe the Lagrangian of the dark neutrino scenario.
The field content of the theory can be found in Table~\ref{tab:field_content}.
\begin{table}[t]
    \centering
    \begin{tabular}{|l|c|c|c|}\hline\hline
        Field & $SU(2)_L$ & $U(1)_Y$ & $U(1)_D$ \\ \hline
        $N_R$   &  1  &  0  &  +1  \\ \hline
        $N_L$   &  1  &  0  &  -1  \\ \hline
        $H_1$   &  2  &  0  &  +1  \\ \hline
        $\phi_1$&  1  &  0  &  +1  \\ \hline
        $\phi_2$&  1  &  0  &  +2  \\ \hline \hline
    \end{tabular}
    \caption{Field content of the dark neutrino scenario. The standard model is not charged under the new symmetry $U(1)_D$.}
    \label{tab:field_content}
\end{table}
The mass Lagrangian is
\begin{equation}
    \mathcal{L}_{\rm mass} = y_\nu \overline{L} \tilde H_1 N_R + M \overline{N_L} N_R + y_D \phi^*_2 \overline{N_R^{c}}N_R
    + y'_D \phi_2 \overline{N_L^{c}}N_L,
\end{equation}
where we denote the left- and right-handed dark neutrino fields by $N_L$ and $N_R$.
After $U(1)_D$ breaking, this gives rise to an inverse seesaw texture
\begin{equation}
  M_\nu = \left(\begin{matrix}
  0 & y_\nu \langle H_1 \rangle & 0 \\
  y_\nu^\dagger \langle H_1 \rangle & y_D \langle \phi_2 \rangle & M \\
  0 & M & y_D' \langle \phi_2 \rangle
  \end{matrix}\right),
\end{equation}
in the basis $(\nu_L, N_R^c, N_L)$.
Note that we have taken the vacuum expectation values (vevs) of the new scalar fields to be real.
This leads to active neutrino masses of the order 
\begin{equation}
    m_\nu\sim y_D' \langle \phi_2\rangle \frac{y_\nu^2 \langle H_1\rangle^2}{M^2},
\end{equation}
as long as $y'_D\langle\phi_2\rangle\ll y_\nu\langle H_1\rangle\ll M$,
while the mixing between active neutrinos and $N_R^c$ is given by $\theta\sim y_\nu \langle H_1\rangle/M$, effectively decorrelating mixing from masses, as is usual in inverse seesaw scenarios.
The $\phi_1$ field is necessary to induce small vevs to both $H_1$ and $\phi_2$ (see Ref.~\cite{Bertuzzo:2018ftf} for details).

From a phenomenological perspective, the mixing among flavor eigenstates gives rise to a coupling between the weak gauge bosons and the neutrino mass eigenstates, namely
\begin{equation}
    \mathcal{L}_{\rm int} = \frac{g}{\sqrt{2}} \sum_{\alpha}^{e,\mu,\tau} \sum_{i=1}^n U_{\alpha i} \overline{\ell}_\alpha \gamma_\mu P_L \nu_i W^\mu + {\rm h.c.} + \frac{g}{c_W}\sum_{\alpha}^{e,\mu,\tau} \sum_{i=1}^n \sum_{j=1}^n U_{\alpha i}^* U_{\alpha j} \overline{\nu}_i \gamma_\mu P_L\overline{\nu}_j Z^\mu,
\end{equation}
where $g$ is the weak coupling constant, $c_W$ is the cosine of the weak mixing angle, and $U_{\alpha i}$ is the $3\times n$ PMNS matrix, where $n$ is the total number of light and heavy neutrino states.
For simplicity, we will assume that the PMNS matrix is standard except for the inclusion of $U_{\mu 4}$.
Denoting the heavy neutrino mass eigenstate as $N_D$, the branching ratio of the $W$ boson to the new decay mode becomes
\begin{equation}
    \mathcal{BR}(W\to \mu N_D) \simeq |U_{\mu4}|^2\mathcal{BR}(W\to \mu \nu),
\end{equation}
assuming $|U_{\mu4}|^2\ll 1$ and $m_{N_D} \ll m_W$.
The heavy neutrino then would decay $N_D\to Z_D \nu$ via the same mixing $U_{\mu4}$, followed by $Z_D\to e^+ e^-$ due to the kinetic mixing with the photon whose strength is given by $\epsilon$.
The width of $N_D$ is given by, in the limit $m_{Z_D}\ll m_{N_D}$,
\begin{equation}
    \Gamma(N_D\to\nu Z_D) \simeq \frac{g_D^2 |U_{\mu4}|^2}{8\pi}\frac{m_{N_D}^3}{m_{Z_D}^2} 
    = \frac{g_D^2}{45~{\rm nm}}\left(\frac{|U_{\mu4}|^2}{10^{-7}}\right)\left(\frac{m_{N_D}}{100~{\rm MeV}}\right)\left(\frac{30~{\rm MeV}}{m_{Z_D}}\right)^2.
\end{equation}
In this regime, for the parameters of interest, the $N_D$ decay can be always considered prompt.
For $Z_D$, on the other hand, the width depends on kinetic mixing with the photon, originating from the term
\begin{equation}
    \mathcal{L}_{\rm km}=\frac{\epsilon}{2} F^{\mu\nu}F'_{\mu\nu},
\end{equation}
where $F$ and $F'$ are the photon and \zdark\ field strengths, and $\epsilon$ is the kinetic mixing parameter.
The partial width to $e^+e^-$ is, in the limit $m_e\ll m_{Z_D}$,
\begin{equation}
    \Gamma(Z_D\to e^+e^-)\simeq\frac{\alpha \epsilon^2}{3}m_{Z_D}
    = \frac{1}{0.1~{\rm mm}}\left(\frac{\alpha\epsilon^2}{2\times10^{-10}}\right)\left(\frac{m_{Z_D}}{30~{\rm MeV}}\right),
\end{equation}
where $\alpha$ is the fine structure constant.
For the parameters of interest, the decay of $Z_D$ happens within the inner tracker in nearly all cases.
While a heavy \zdark\ would generally decay promptly, lighter bosons would experience larger Lorentz boosts from the $W$ decay leading to macroscopic displacements (e.g. 1\,mm to 10\,cm) and thus a displaced vertex signature.
For a heavier $Z_D$, other decay channels will open up, and the partial widths can be properly obtained using the $R(s)$ ratio, that is, the ratio between the cross sections for $e^+e^-$ to hadrons versus muons, as in Ref.~\cite{Curtin:2014cca}.
We have implemented the dark neutrino model in FeynRules~\cite{Christensen:2008py, Christensen:2009jx} and created the set of UFO~\cite{Degrande:2011ua, Darme:2023jdn} files
needed for use within BSM event generator tools. The \texttt{DarkNeutrino} model is made publicly available on Zenodo~\cite{herwig_christian_2023_8277735}.

\section{LHC Search strategy}
\label{sec:concept}

The dark neutrino model described above opens two potential portals to produce dark sector particles from their SM counterparts, through either mixing of the neutral leptons or $U(1)$ gauge bosons.
This suggests the potential for multiple search strategies to contribute complementary sensitivity, including experiments that employ a range of particle species and energies.
In addition to the high-intensity neutrino sources like the FNAL Booster Neutrino Beam that serviced the MiniBooNE experiment, beam-dump and fixed-target setups can similarly explore this model through neutrinos produced in light mesons and muon decays~\cite{Atre:2009rg,deGouvea:2015euy}.
Alternatively, the $U(1)_D$ portal can be explored through a range of experiments targeting dark photon production and prompt decay, including $e^+e^-$ colliding beams, fixed targets, and meson decays.
High-energy proton-proton collisions can also present powerful constraints on this possibility at high mass~\cite{LHCb:2019vmc}.
Specific constraints depend on the dominant portal coupling that may be realized in Nature, in addition to the masses of the BSM states.
For the range of parameters considered in this work, the dark neutrino decays promptly, so that constraints on long-lived heavy neutral leptons do not apply.

\begin{figure}[t!]
\centering
\includegraphics[height=0.22\textwidth]{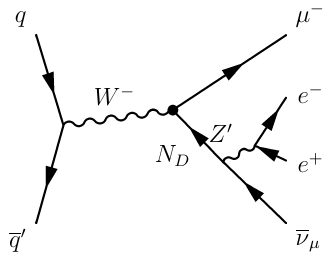}
\includegraphics[height=0.22\textwidth]{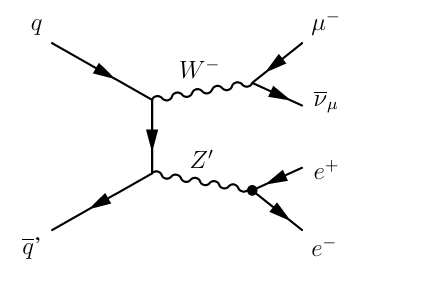}
\includegraphics[height=0.22\textwidth]{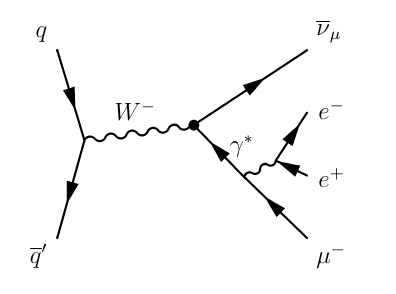}
\caption{Representative diagrams are shown for the production of $e^+e^-$ pairs from \zdark\ decays in $W^-\to\mu^-\bar\nu_\mu$ events, produced either through the mixing of neutral leptons (left) or $U(1)$ gauge forces (center). At right the dominant SM background process is shown, in which the muon radiates an $e^+e^-$ pair via virtual photon emission.}
\label{fig:feynW}
\end{figure}

The LHC provides a large dataset of neutrinos across a range of energies that may be exploited to test scenarios across a range of values for the neutral lepton mixing parameter \UmFour.
At the ATLAS and CMS experiments, neutrinos from decays of the massive weak bosons provide a promising source, most notably the $W^\pm$.
Leptonic decays $W^\pm\to\ell^\pm\nu$ ($\ell=e$ or $\mu$) present a promising experimental channel signature because of the presence of an energetic charged lepton that can be used to trigger and cleanly reconstruct the event.
Figure~\ref{fig:feynW} (left) shows a diagram corresponding to signal production through this process, where a dark neutrino is produced and decays to two charged leptons and a SM neutrino.
The same final state can access the \zdark\ portal through diagrams such as the one shown in Figure~\ref{fig:feynW} (center), though it is generally subdominant for the combination of model parameters considered in this work.
Figure~\ref{fig:feynW} (right) shows a representative diagram for the dominant SM background process to this signature, where a virtual photon is radiated off of the charged lepton from the $W$ decay.

In the following sections, the potential sensitivity of this channel is investigated using Monte Carlo simulation to compare the dark neutrino and SM processes and devise a promising search strategy.
The key feature to be targeted is the low-mass pair of charged leptons produced through the \zdark\ decay.
This resonance provides a clean experimental signature and allows a robust ``bump hunt'' strategy to extract the peaking signal over potential background processes though the use of sideband techniques.
For light \zdark\ masses, the macroscopic displacement of the lepton pair can also be used as an effective method of background rejection.

Because the \zdark\ decays via coupling to SM hypercharge, the spectrum of possible final states depends only on its mass.
The branching fraction to leptons generally increases for shrinking \zdark\ masses, with significant exceptions near the hadronic resonances that can be calculated using $R(s)$ ratio data~\cite{Curtin:2014cca}.
As an example, $\mathcal{BR}(\zdark\to e^+ e^-)\sim 15$\% at 10 GeV, growing to nearly 50\% at 400\,MeV.
Below this value, the fraction of decays to the electron and muon channels begin to diverge, with $\mathcal{BR}(\zdark\to e^+ e^-)=100$\% below $2m_{\mu}$.
Unless otherwise specified we take the sum of the electron and muon channels as the signal in the remainder of this work.

\section{Simulated samples and detector-level reconstruction} 
\label{sec:mc}

Simulated samples of Monte Carlo events corresponding to SM processes and the dark neutrino signals are used to study analysis selection criteria and establish sensitivity to the BSM process.
SM backgrounds are calculated using \textsc{MG5\_aMC@NLO} 3.4.1 at LO~\cite{Alwall:2014hca,Alwall:2014bza}, interfaced to \textsc{Pythia} 8.306~\cite{Bierlich:2022pfr}, including the dominant process of inclusive $\ell^\pm \bar\nu_\ell \ell'^+ \ell'^-$ production. 
Leptons are generated with $\pt>1$\,GeV and $|\eta|<5$, and requiring that the mass of the same-flavor, opposite-sign dilepton pair $m_{ee}$ is greater than 10 MeV. 
In the inclusive phase space, the leading contribution to this process comes when a virtual photon is radiated off the charged lepton resulting from the $W$ decay, with sub-leading by radiation off of an initial-state quark, shown in Figure~\ref{fig:feynW}(right).
The process is normalized by applying a $k$-factor corresponding to the full NNLO + NLO electroweak charged-current prediction~\cite{Anastasiou:2003ds}.
The corresponding process with a $Z$ boson produced in association with a pair of low-mass electrons in place of a $W$ is also simulated at LO and normalized to the NNLO calculation~\cite{PhysRevD.51.44,HAMBERG1991343,VANNEERVEN199211} in a similar manner.
Lastly, a MC calculation corresponding to the production of top-quark pairs in association with a soft virtual $Z/\gamma\to\ell^+\ell^-$ is used, modified by the ratio of NNLO+NNLL to LO cross section for top-quark pairs~\cite{Cacciari:2011hy,Czakon:2013goa}.
The expected contribution from rarer SM processes yielding three or more prompt leptons is negligible for the phase space considered in this analysis.

Signal samples are generated with the same tools, using the Dirac \textsc{DarkNeutrino} model introduced in Section~\ref{sec:theory}.
Our results depend only weakly on the Dirac versus Majorana nature of the dark neutrino.
Dark neutrinos produced via $W^\pm$ decay are simulated from the $\ell^- \bar\nu_\ell \zdark(\to\ell\ell)$ matrix element.
As a baseline selection of parameters, we set $|\UmFour|^2=10^{-4}$, $\alpha\epsilon^2=2\times10^{-10}$, and $g_D=0.25$, and consider masses $m_{\zdark}<m_{\ndark}$ corresponding to the prompt 2-body decay of the $\ndark$.
In this configuration
 $\ell^- \bar\nu_\ell \zdark$ production is dominated by the neutrino portal rather than direct radiation of a \zdark\ off an initial-state quark (though the latter can become important for very small values of $|\UmFour|^2$) and the cross section is largely insensitive to $\epsilon$ and $g_D$.
To account for higher-order effects that are not included in the signal calculation, the $k$-factor for the SM $W^\pm$ process is also applied to signal.
At $\sqrt{s}=13\Tev$ the resulting cross section is
$2.1\pb$ 
for when $m_{\ndark}=300\Mev$ and $m_{\zdark}=100\Mev$, but varies within less than 10\% when $1\Mev<m_{\zdark}<m_{\ndark}<10\Gev$.

Simulated particle-level quantities are translated to detector-level observables with parameterized smearing functions in order to estimate the potential reach of the ATLAS and CMS experiments to the signatures above.
Because the key characteristics of the signal are encoded in the kinematics of the \zdark, we restrict ourselves to the consideration of leptonic observables only in this work.
While improved sensitivity may be achieved by considering other features of the collision events (such as the imbalance of the transverse momenta of all reconstructed particles), these can depend  strongly on the details of the experimental resolution and running conditions, and are thus omitted from the present study.

\begin{figure}[t!]
\centering
\includegraphics[width=0.45\textwidth]{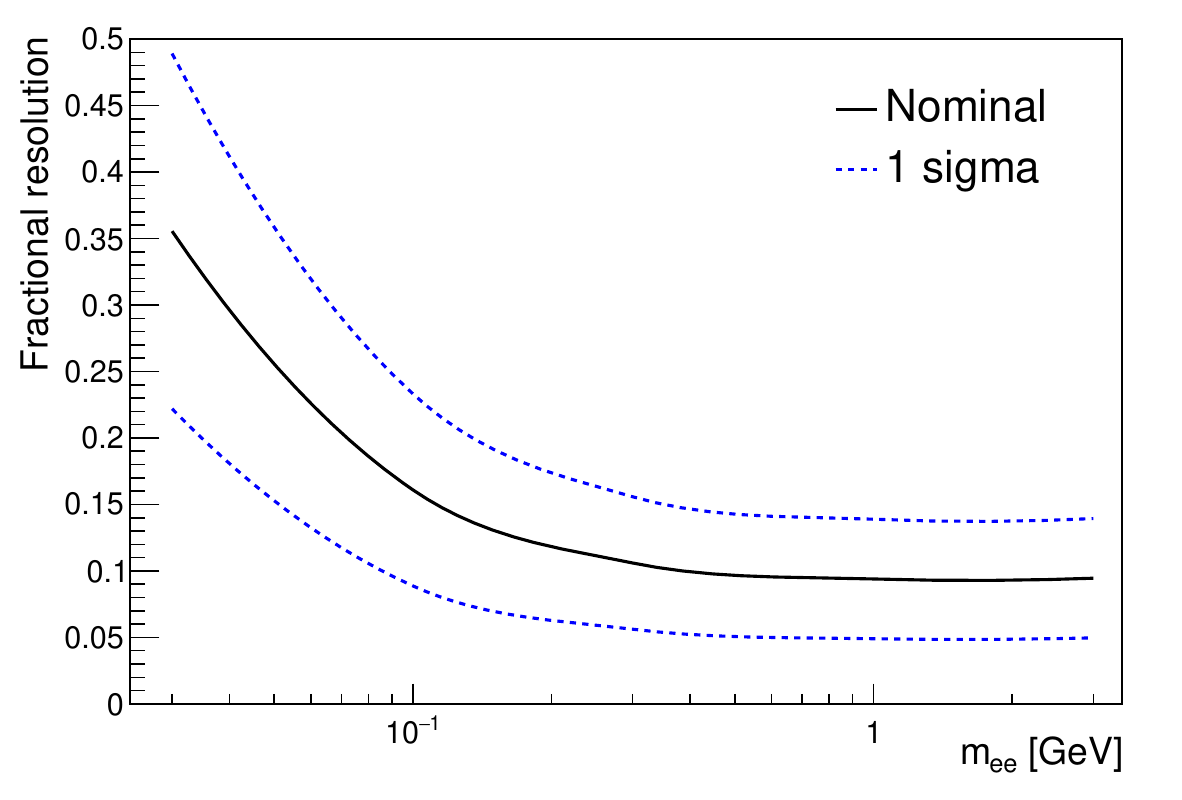}
\caption{The di-electron mass resolution is shown for electrons for the optimistic and pessimistic scenarios described in the text. 
Single-electron resolutions are applied to simulated events in the dark neutrino signal model for a 10\Gev\ \ndark\ mass.
The quoted resolution is computed as the smallest interval containing 68\% of the smeared mass values.}
\label{fig:massrel}
\end{figure}

High-energy dark photons produced through the signal processes described above lead to pairs of high-energy particles that must be exceptionally collimated in the case of a low-mass \zdark.
Consequently, the resolution on the \zdark\ dilepton mass $m_{\ell\ell}$, which drives the background rate, relies on the precision of the leptons' angular reconstruction.
We apply a \pt-dependent Gaussian smearing on reconstructed electron kinematics, corresponding to 0.001-0.0025 radians in $\phi$ and 0.0005-0.002 in units of $\cot\theta$.
In addition, a 2.5-8\% smearing on the electron \pt\ is assessed, taken as a conservative estimate in the phase space most relevant to this search~\cite{CMS:2014pgm,CMS:2020uim}.%
\footnote{We adopt standard collider reference coordinate system centered at the interaction point with the $z$-axis oriented along the beamline, the $x$-axis pointing to the center of the LHC ring, and the $y$-axis pointing upward. The $(r,\phi)$ plane is transverse to the beam, with $\theta$ describing the polar angle from $+\hat{z}$.}
Figure~\ref{fig:massrel} shows the mass resolution that is obtained for pairs of electrons with these smearing functions applied, as evaluated on signal MC samples with \ndark\ mass fixed to 10\Gev.
The variation found by considering the upper and lower extremes of the single-object kinematic resolutions quoted above correspond to the region enclosed in the $\pm1\sigma$ band.
In the muon decay channel, the mass resolution is well-established from measurements of light meson decays, which we conservatively take as 2\%.
A 15\% inefficiency factor per lepton is also applied to account for the effect of requirements on the reconstruction quality entering through identification and isolation cuts.
While most events with four leptons within the detector acceptance described above should be removable via loose lepton or track vetos, we assume a 5\% inefficiency factor that allows these events to populate our signal region.

The lab-frame displacement of the \zdark\ is calculated for signal events, based on the proper lifetime (proportional to $1/\epsilon^2 m_{\zdark}$) and large momentum inherited from its production in the $W$ boson decay.
For the analysis considering SM backgrounds with prompt leptons, signal events are required to have displacements less than 1~\,mm.
For signal events with displacements from 1~\,mm to 10~\,cm, the SM backgrounds described above can be efficiently rejected based on the separation of the dilepton vertex from the primary vertex, which can be well-reconstructed from the high-\pt\ $W$-decay lepton and the recoil system.
In this case it is reasonable to expect that a background-free search can be conducted, at the cost of some additional penalty of the signal efficiency described in Section~\ref{sec:analysisDisp}.

\section{Analysis of dark neutrino production in \texorpdfstring{$W$}{W} boson decays}
\label{sec:analysisW}

To maximize the sensitivity to small neutrino portal couplings, we aim to construct a relatively inclusive analysis retaining high signal efficiency.
Thus, typical $W$ boson progenitors of the BSM signal will be produced nearly at rest, leading to a single high-\pt\ lepton in one hemisphere of the event opposite a low-mass, low-momentum di-lepton pair in the other.
The dominant background process is $W$ boson production involving the radiation of a virtual $Z/\gamma^*\to\ell^+\ell^-$.
This can originate from `diboson-like' processes, where the lepton pair is radiated from an initial-state quark or the $W$ propagator, or final-state radiation off the charged lepton produced in the $W$ decay.
Before any selection is applied, the largest contribution comes from the latter class of processes, which can be effectively reduced by requiring the soft di-lepton pair to be well-separated from the third, high-\pt\ lepton.

Rarer backgrounds arise from analogous processes where a soft lepton pair is produced in association with a $Z$ boson or a top and anti-top quark pair.
A challenging source of background can arise from $Z$ to four lepton events, when one of the $Z\to\ell^+\ell^-$ daughter leptons radiates another soft lepton pair
if the high-\pt\ radiating daughter lepton fails to be reconstructed.
While backgrounds with top quarks can lead to a similar final state, they can be suppressed by the  accompaniment of hadronic activity. 
Other processes are either significantly rarer than those considered above or can be well-suppressed by similar requirements.
Experimental backgrounds may also arise from the mis-identification of charged hadrons and non-prompt leptons and must be estimated directly from the data.
Uncertainties stemming from this and other sources are discussed further in Section~\ref{sec:stats}.

\subsection{Event selection}

A baseline set of events is defined for the analysis which should ensure the signal can be well-reconstructed by the experimental apparatus.
Three leptons are required, each of which enters the central region of the detector ($|\eta|<2.5$) and meets a minimum \pt\ requirement.
These are kept as low as possible to maximize the signal rate from the light \zdark\ decay, taking 3\Gev\ muons and 5\Gev\ electrons as the baseline.
A scenario including electrons with $\pt>1\Gev$ is also considered, motivated by more aggressive strategies being pursued by the CMS Experiment~\cite{CMS-DP-2019-043}.
The majority of $W\to\ell\nu$ decay events can be collected through the use of triggers requiring a single high-\pt\ electron or muon at ATLAS and CMS. 
Thus, at least one of the three well-reconstructed leptons should also satisfy $\pt>25\Gev$, which is taken as a representative choice for the suite of trigger paths used across experiments and lepton flavors.
The mass of the two leptons with the lowest transverse momenta \mll\ is required to be less than 10 GeV to focus on the region most consistent with the MiniBoone anomaly.

As the signal is a purely electroweak process, events with additional hadronic activity such as top quarks decays are vetoed.
The sum of charged hadronic momenta $H_T$ is defined as the \pt\ sum of all charged hadrons satisfying $\pt>2\Gev$ and $|\eta|<2.5$ and is required to be less than 30\Gev.

Figure~\ref{fig:presel} shows a comparison of various signals and the cumulative SM background processes in the electron channel, after these baseline requirements are enforced.
The expected yields for $W$, $Z$, and top-quark processes are shown for 13\Tev\ proton-proton center of mass energy, normalized to 150\ifb\ of data, corresponding to the approximate sizes of the Run~2 data sets collected by each of the ATLAS and CMS Experiments from 2015-2018.
The expected spectrum of events is also shown for the dark neutrino model for several choices of \zdark\ and \ndark\ masses.
Leptons are labeled $l_1, l_2, l_3$ in order of descending \pt.

\begin{figure}[t!]
\centering
\includegraphics[width=0.45\textwidth]{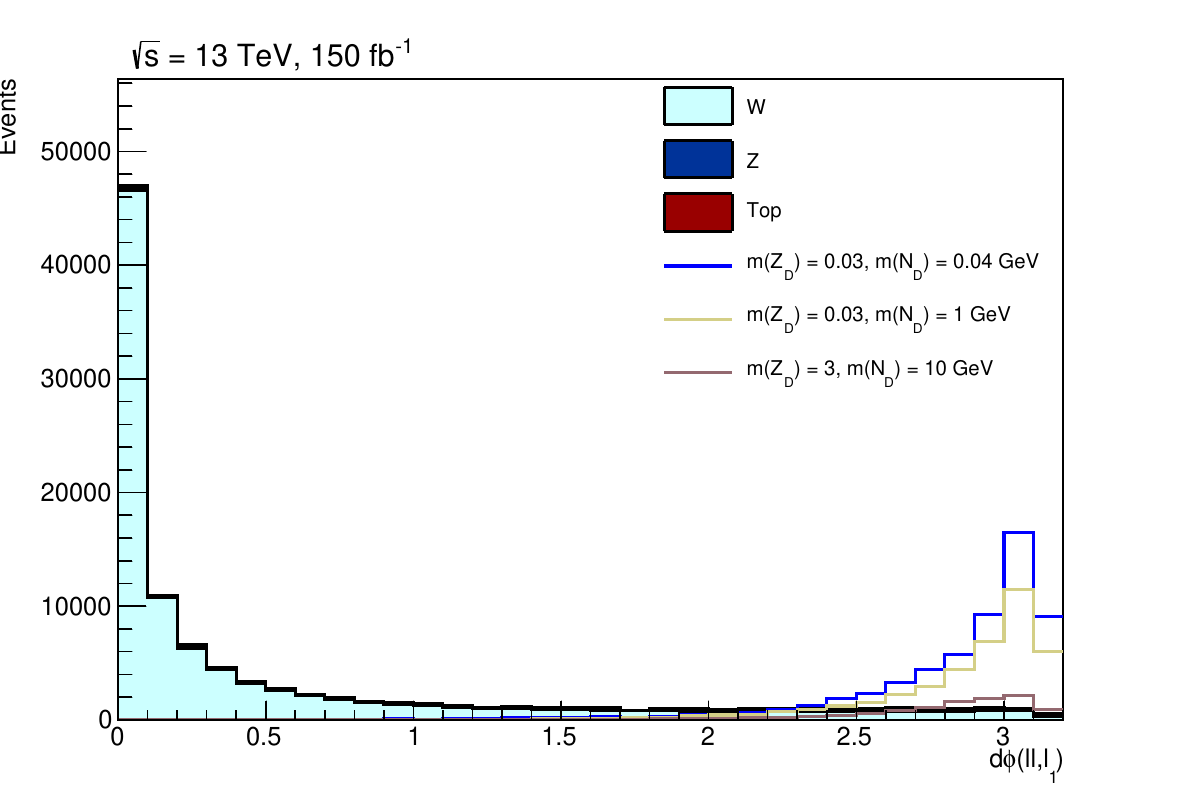}
\includegraphics[width=0.45\textwidth]{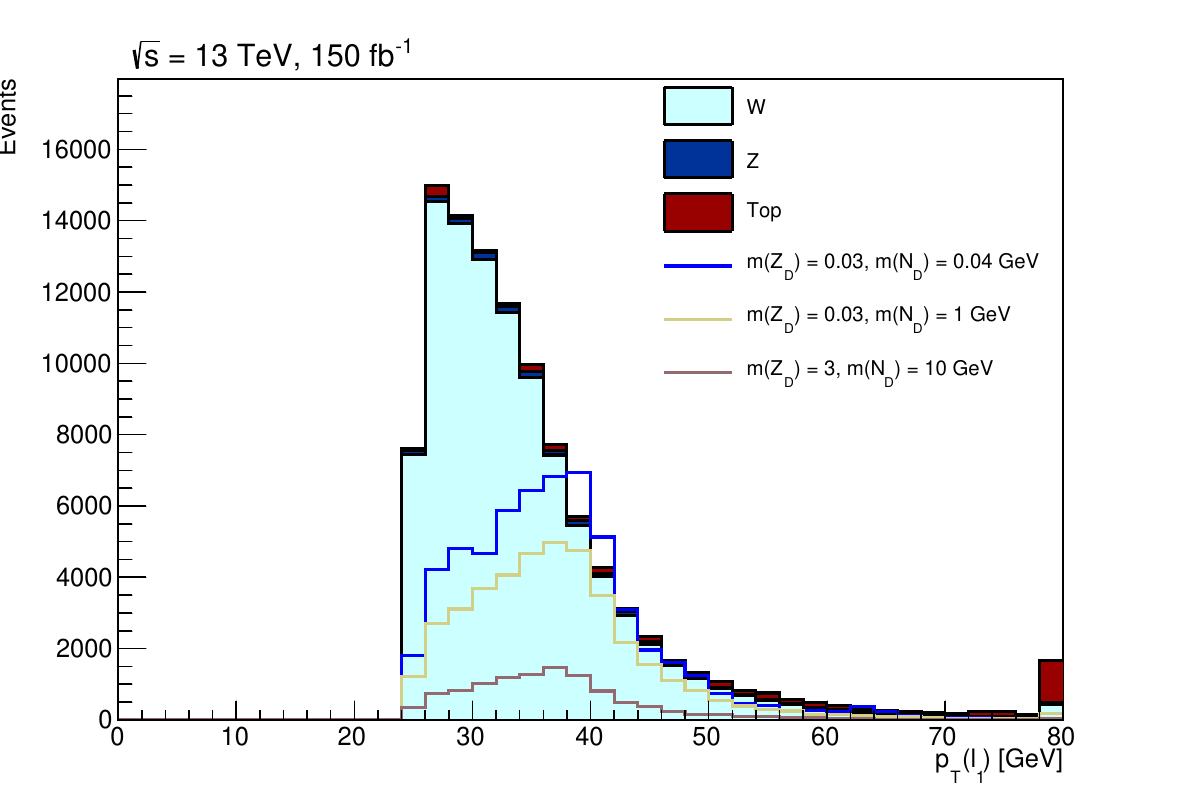}
\includegraphics[width=0.45\textwidth]{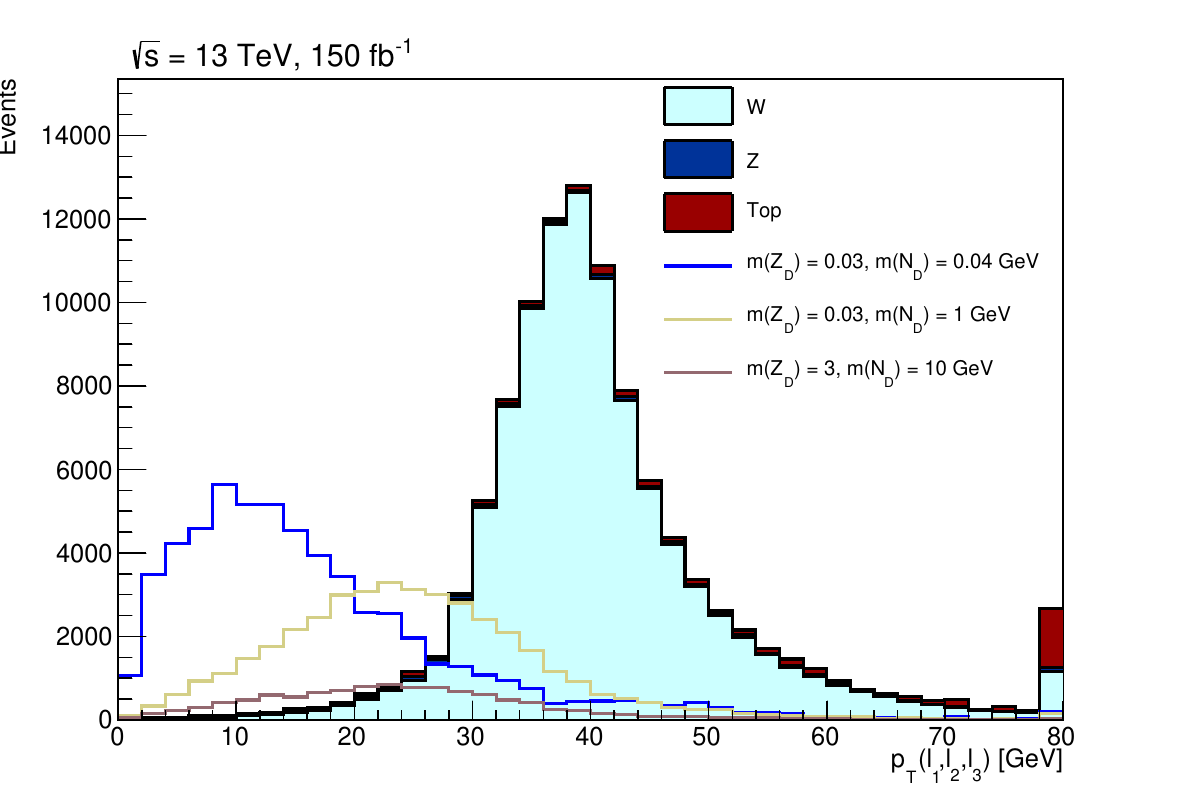}
\includegraphics[width=0.45\textwidth]{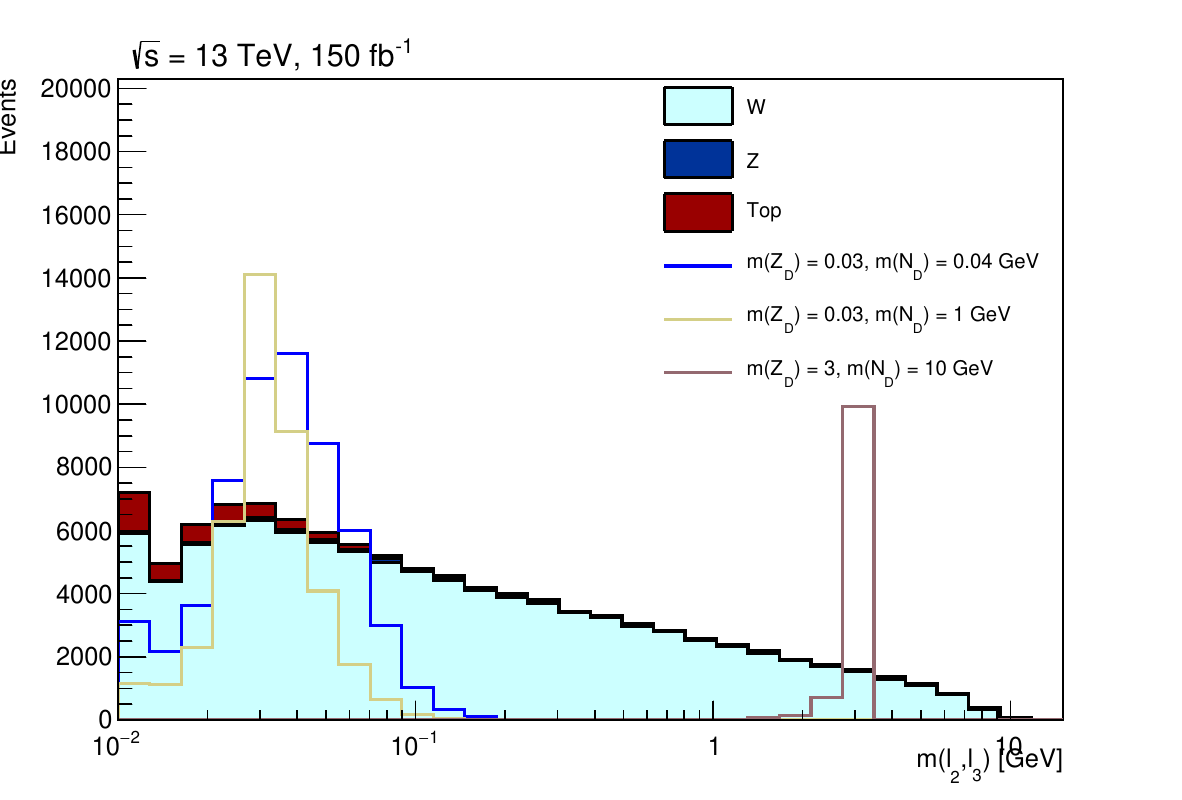}
\caption{The expected distribution of BSM signal and background processes are shown for events passing the pre-selection criteria, normalized to 150\ifb\ of collected data.
SM background contributions are shown cumulatively for $W$, $Z$, and top processes, while realizations of the signal model for different \zdark\ and \ndark\ masses are overlaid, fixing $|\UmFour|^2=10^{-4}$.
The electron channel is shown for events with lepton $\pt>1$~GeV.
Overflow values falling below (above) the x-axis limits are included in the first (last) bin.
}
\label{fig:presel}
\end{figure}

Several key differences between the signal and BSM processes can be exploited to define a set of events enriched in potential dark neutrino decays.
In the majority of SM $W$ decay events, the soft di-lepton pair is most often radiated off of the charged lepton, as opposed to off of a neutrino in the BSM process.
In addition to the large angular difference $d\phi(\ell_1,\ell_2\ell_3)$ between the highest-\pt\ lepton and combined system of the remaining two leptons, this lowers the average momentum of the charged lepton (neutrino) from the initial $W$ decay in background (signal) events.
As a consequence, signal events pass the high-\pt\ lepton trigger more efficiently and lead to a smaller imbalance in transverse momentum among the reconstructed visible objects (\ptmiss).
The last effect can also clearly be seen in the magnitude of the vector sum of the three lepton momenta $p_{T}(\ell_{1},\ell_{2},\ell_{3})$, equivalent to \ptmiss\ in the limit where the $W$ boson is produced at rest.
The mass spectra of the the soft di-lepton pair is smoothly-falling for the SM background.

Table~\ref{tab:cuts} outlines a set of selection criteria based on these differences, designed to significantly remove the background while retaining a high signal efficiency for all combinations of \ndark\ and \zdark\ masses.
The azimuthal difference between the leading lepton and the \zdark\ candidate formed by the remaining two leptons must be larger than 2.7 radians.
Additionally, the distance $dR=\sqrt{d\phi^2+d\eta^2}$ between the leading lepton and \zdark\ candidate is required to be between 2.7 and 3.5.
The mass of the three leptons is required to be less than 80.3\Gev\ to be consistent with the $W$ decay hypothesis.
The \pt\ of the three-lepton system $p_{T}(\ell_{1},\ell_{2},\ell_{3})$ should be less than half the $W$ mass.

\begin{table}[t!]
    \begin{center}
    \begin{tabular}{c|c}
        Observeable & Selection  \\
        \hline
        $\ell$ multiplicity & exactly three \\
         Electron $p_{T}$ & $>1, 5$ GeV   \\
         Muon $p_{T}$ & $>3$ GeV   \\
         $|\eta_{\ell}|$ & $<2.5$   \\
         $p_{T}(\ell_1)$ & $>25$ GeV   \\
         \hline
         $d\phi(\ell_{1},\zdark)$ & $>2.7$ \\
         $dR(\ell_{1},\zdark)$ & $\in[2.7,3.5]$ \\
         $m(\ell_{1},\ell_{2},\ell_{3})$ & $<m_W$  \\
         $p_{T}(\ell_{1},\ell_{2},\ell_{3})$ & $<m_W/2$  \\
    \end{tabular}
        \caption{Summary of the selection criteria, including baseline cuts in addition to selections optimized for the LHC Run~2 data set. Scenarios with a minimum electron \pt\ of both 1 and 5\Gev\ are considered.}
        \label{tab:cuts}
    \end{center}
\end{table}

Figure~\ref{fig:cutsLoose} shows several properties of the remaining events that pass this selection, where we have adjusted the mixing to $|\UmFour|^2=10^{-6}$ for visual clarity.
At this point several key handles remain to discriminate signal from the SM background.
While the mass of the di-lepton system is smoothly falling for the background, the signal process peaks at the \zdark\ mass.
The value of the \ndark\ mass also significantly impacts the signal kinematics.
When the \ndark\ is significantly heavier than the \zdark\ the lepton \pt\ spectra tends to peak at low values.
When the mass difference between the new states becomes small, the majority of the \ndark\ momenta is transferred to the leptons through the decay $\ndark\to\nu(\zdark\to\ell^+\ell^-)$.
As a consequence, less momentum is carried by the neutrino, the charged lepton \pt\ favor larger values, and the three-lepton mass peaks closer to $m_W$.
While our projected limits are based on the reconstructed \zdark\ candidate mass alone, future work may exploit this information using more sophisticated techniques such as multivatiate discriminants, kinematic fits, or matrix element methods to take advantage of these features.

\begin{figure}[t!]
\centering
\includegraphics[width=0.45\textwidth]{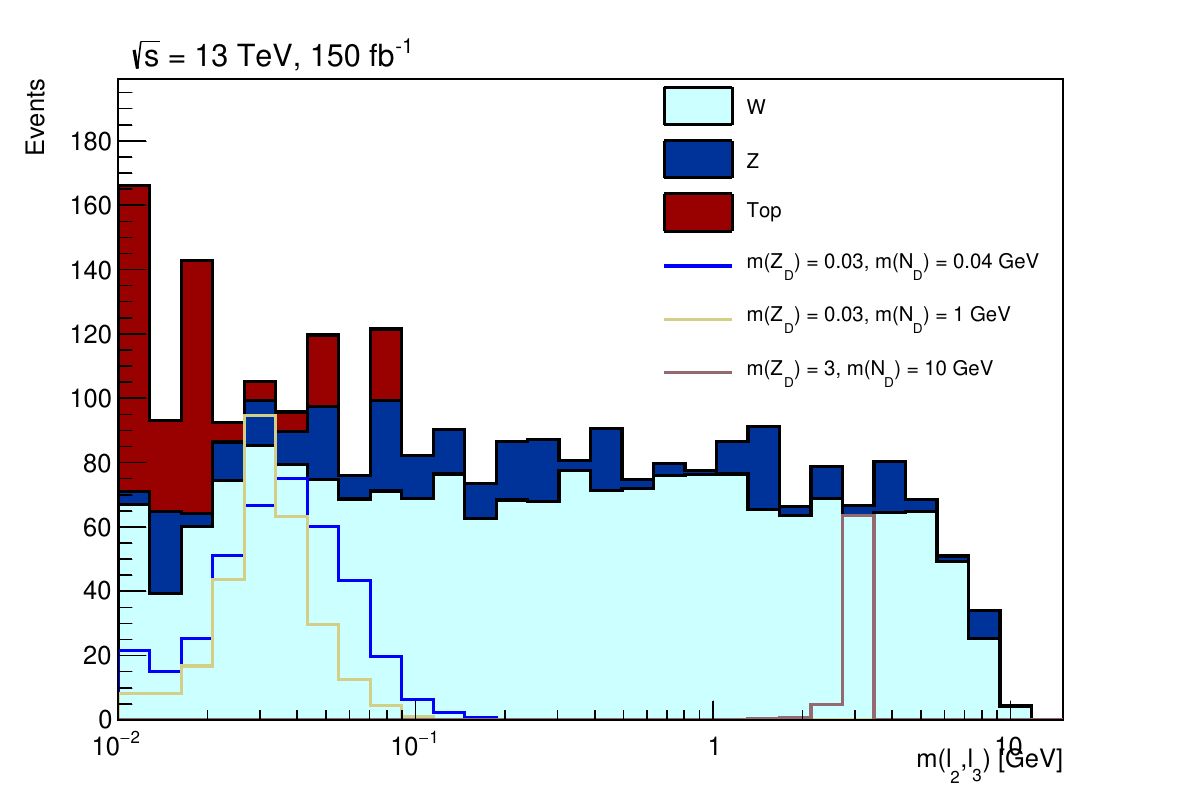}
\includegraphics[width=0.45\textwidth]{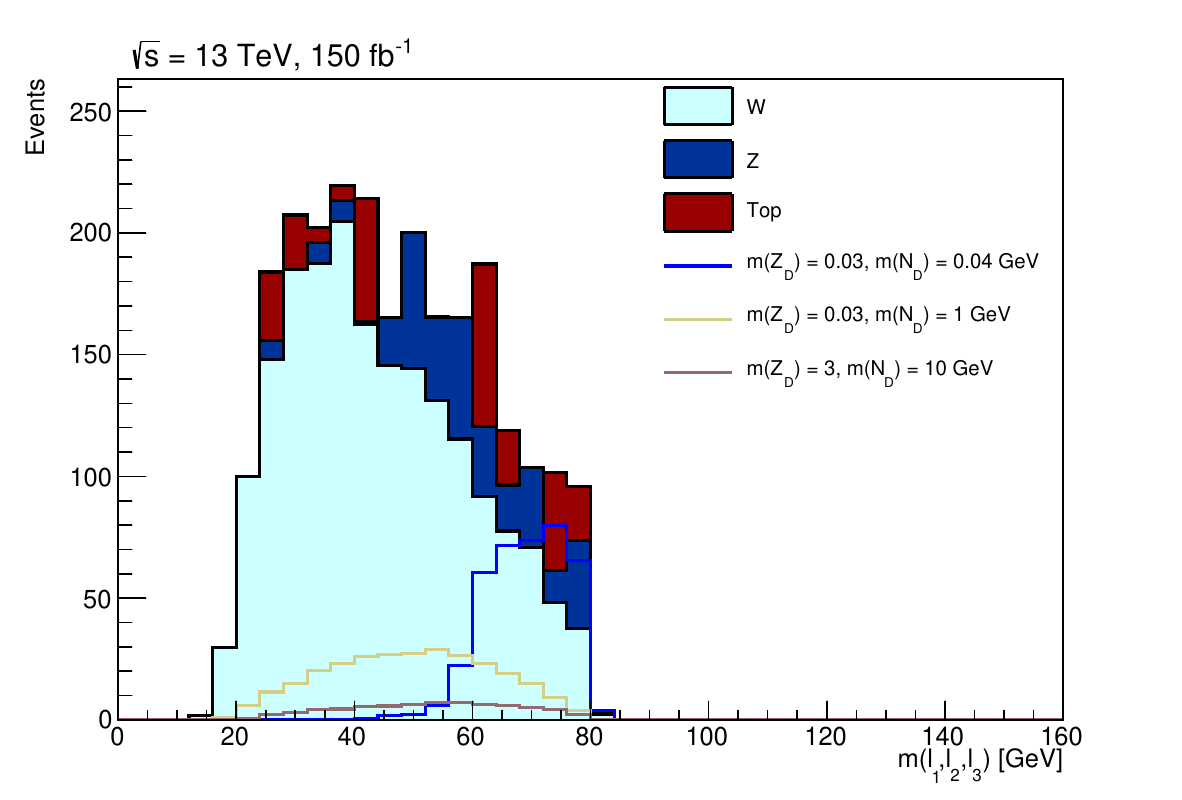}
\includegraphics[width=0.45\textwidth]{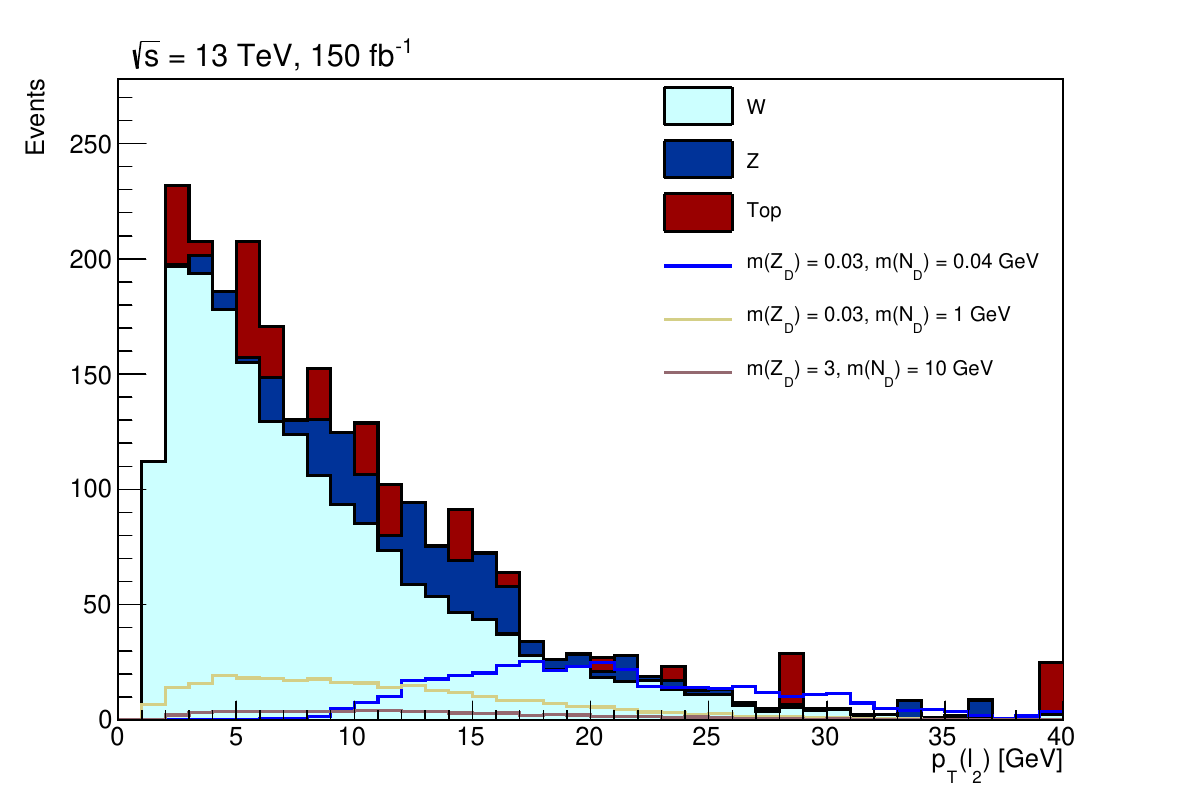}
\includegraphics[width=0.45\textwidth]{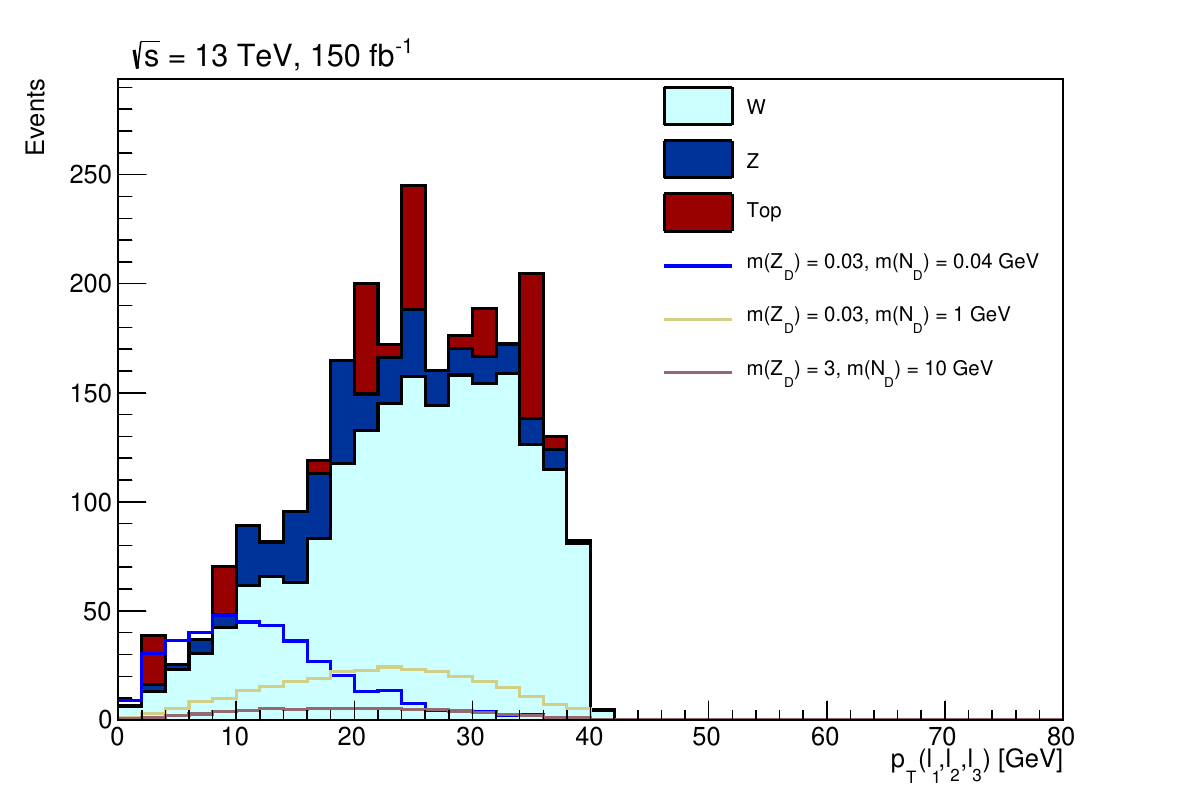}
\caption{
The expected distribution of BSM signal and background processes are shown for events passing the signal region selection, normalized to 150\ifb\ of collected data.
SM background contributions are shown cumulatively for $W$, $Z$, and top processes, while realizations of the signal model for different \zdark\ and \ndark\ masses are overlaid, fixing $|\UmFour|^2=10^{-6}$.
The electron channel is shown for events with lepton $\pt>1$~GeV.
Overflow values falling below (above) the x-axis limits are included in the first (last) bin.
}
\label{fig:cutsLoose}
\end{figure}

\subsection{Signal extraction procedure and uncertainties}
\label{sec:stats}

The mass of the lepton pair from the \zdark\ decay has not been explicitly included in the selection described above, and can be used to extract signals in the presence of the smoothly falling SM background distribution.
This feature allows the non-resonant background to be estimated from a fit using the `data sideband' technique, whereby interpolation to \mll\ values populated by the signal can commonly constrain the expected background to the sub-\% level.
A significance metric is calculated from the number of expected signal $N_S$ and background events $N_B$ in a di-lepton mass window as $S=N_S/\sqrt{N_B}$, where models that predict $N_S$ corresponding to $S>S_{excl}\equiv2$ are expected to be excluded.
Because the mass resolution varies with both the \zdark\ candidate mass itself as well as the lepton flavor, as described in Section~\ref{sec:mc}, an optimal window in \mll\ is used to extract $N_S$ for each scenario under consideration.
This significance-maximizing interval ranges from 10 to 500\Mev\ for the electron channel and 3 to 100\Mev\ for the muon channel.
In all scenarios considered, no \mll\ window is selected which contains fewer than 10 background events.

Because the production of \ndark\ in the analysis phase space is dominated by $W$ decays, the expected signal cross section is directly proportional to $|\UmFour|^2$.
This allows a limit on the number of signal events to be translated to a limit on \UmFour\ based on the significance $S_{\mu4\text{,ref}}$ obtained with the sample of Monte Carlo events generated with a reference coupling $|U_{\mu4\text{,ref}}|^2$ via the relation
$|U_{\mu4\text{,excl}}|^2 = (S_\text{excl} / S_\text{ref}) \cdot |U_{\mu4\text{,ref}}|^2$.
This extrapolation is verified by additional MC calculations of the cross sections and kinematic distributions with re-scaled coupling parameters.

This statistical procedure is repeated for modifications of the analysis accounting for the impact of various systematic uncertainties.
Without a complete detector simulation it is difficult to assess the importance of backgrounds due to mis-identified and non-prompt leptons, including $B$ decays and photon conversions.
If lepton identification criteria cannot reduce these backgrounds to a negligible level, they may constitute an additional background that varies smoothly in \mll, up to known hadronic resonances that can be masked.
We address this possibility with a conservative, ad-hoc approach of scaling the background mass template by a factor of 1.5 and applying the full size of the correction as an uncertainty.
The mass resolution also has a large impact on the result, as it effectively determines the number of background events overlapping with a given signal peak.
For muons this is taken to be $2\pm1\%$ independent of mass whereas for electrons the smearing functions and their uncertainties are those described in Section~\ref{sec:mc}.
Uncertainties are also considered to account for the efficiency to select and identify leptons, taken to be $85\pm5\%$.

\subsection{Analysis of displaced \texorpdfstring{\zdark}{ZD} decays}
\label{sec:analysisDisp}

In the case where the \zdark\ is light and weakly-enough coupled to become significantly long-lived in the lab frame, a parallel analysis strategy can be pursued to the prompt search described above.
In order to select events that are well-reconstructed and compatible with the Dark Neutrino signal, we consider the same selection requirements developed for the prompt analysis, with the additional requirement that all leptons have $\pt>5$~GeV.
Instead of requiring the pair of lower-\pt\ leptons to originate from a vertex with displacement less than 1~\,mm, distances of 1~\,mm to 10~\,cm are considered.
The distribution of expected \zdark\ displacements is shown in Figure~\ref{fig:disp} for a range of representative signal model parameter sets.
In the case of a high-mass \zdark, nearly all events populate the prompt analysis bin, whereas a wider distribution of lifetimes (extending beyond a  meter in some cases) are found for the 30\,MeV benchmark.
While the selection requirements are found not to significantly affect the shape of the expected displacement, varying values of the kinetic mixing $\epsilon$ have a significant impact.

\begin{figure}[t!]
\centering
\includegraphics[width=0.49\textwidth]{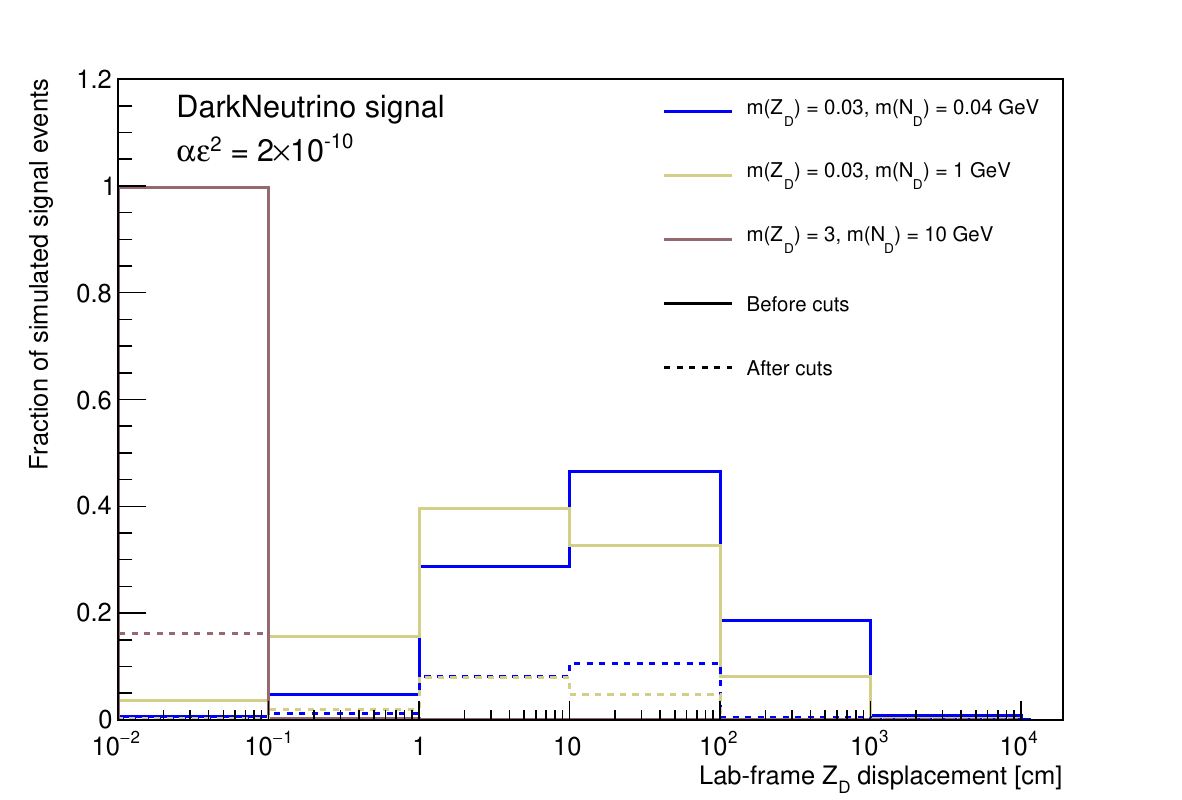}%
\includegraphics[width=0.49\textwidth]{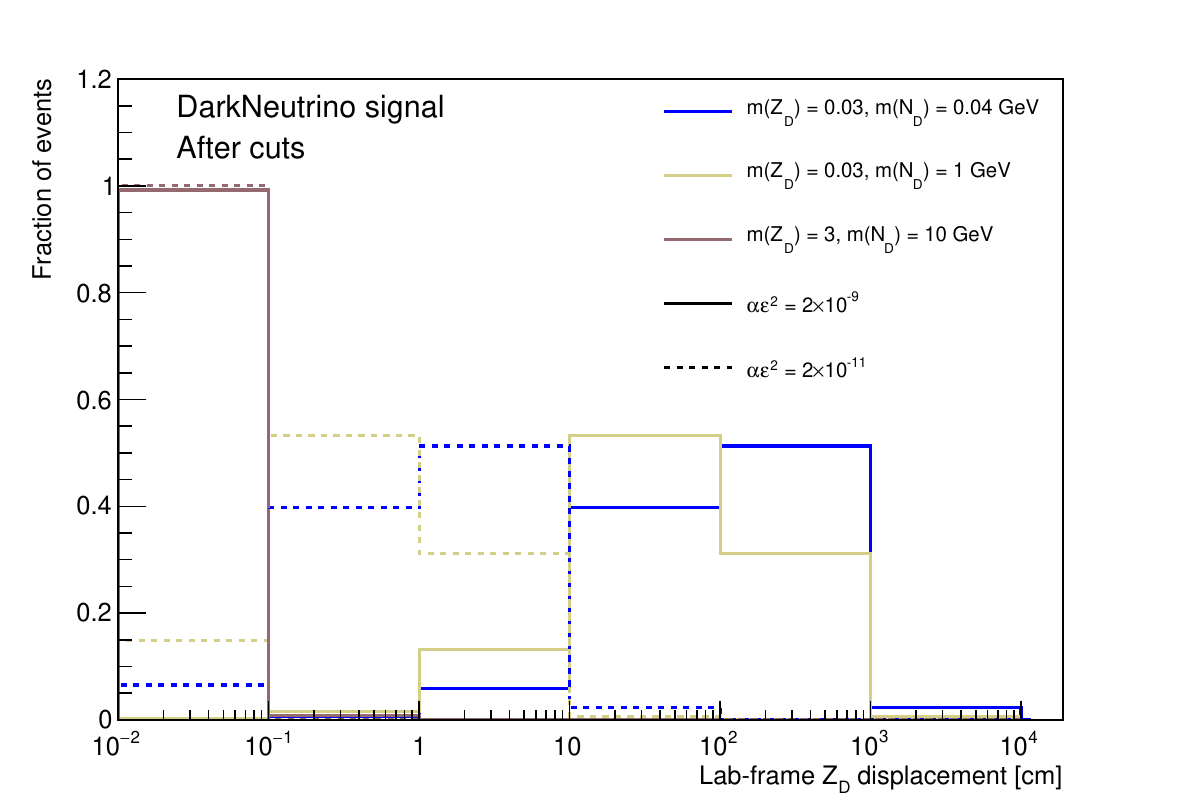}
\caption{
  At left, the distribution of displacements is shown for various combinations of signal model parameters, normalized to unity. The subset of events passing the full event selection is also shown for comparison.  At right, the same distributions are shown for events passing the full event selection with variations of the kinetic mixing parameter $\epsilon$. Events falling below or above of the displayed range of displacements are added to the first or last bin of displacement, respectively.
}
\label{fig:disp}
\end{figure}

Displacements up to 10\,~cm are considered because high lepton reconstruction efficiency should be achievable in this range, with lepton tracks still expected to leave hits in all or nearly all of the silicon tracker layers.
There is no SM process that leads to the expected signature of a pair of resonant leptons with a significant displaced vertex.
Displaced, non-resonant lepton pairs from the decay of B mesons produced in the $W$ recoil may be important for shorter lifetimes.
Unfortunately the estimation of such fake/non-prompt backgrounds are outside the scope of this study, only being reliably estimated using data-driven methods by the LHC experiments.
However analyses studying related signatures have demonstrated that these backgrounds can be effectively controlled at the required level in Run~2 (see, e.g. Refs.~\cite{CMS:2022fut,ATLAS:2022atq}), even before the requirement of a narrow di-lepton mass window.
To estimate the approximate potential of the displaced analysis search strategy, we extract expected limits based on the requirement that at least fifty signal events pass the kinematic requirements.
This value is chosen to allow an additional signal inefficiency of 10\% in order to select a high-quality displaced di-lepton vertex, while still leaving enough expected events to set a limit under the background-free assumption.
Together with the signal efficiency of the kinematic requirements, this total efficiency goal is in line with that reported by similar analyses by ATLAS and CMS~\cite{CMS:2022fut,ATLAS:2022atq}.

\section{Results}
\label{sec:results}

Expected limits are placed on the neutrino portal coupling $|\UmFour|^2$ for a range of signal mass hypotheses and experimental conditions.
Limits are shown for the $\sqrt{s}=13\Tev$ center of mass energy and 150\ifb, corresponding to the potential reach that either ATLAS or CMS might place given data already recorded.
These results are also extrapolated to a potential HL-LHC scenario where 4\iab\ are collected at a $\sqrt{s}=14\Tev$ center of mass energy.
Because this difference in energy is relatively minor, the same sets of MC events are used, with their respective cross-sections scaled to their values at $\sqrt{s}=14\Tev$.

Figure~\ref{fig:limitNominal1} shows projected exclusions for the case of a \zdark\ mass fixed to 30\Mev, varying $m_{\ndark}$.
Existing constraints are also displayed~\cite{Atre:2009rg,deGouvea:2015euy} in addition to the region of parameters favored by Ref.~\cite{Bertuzzo:2018itn}.
Results are shown for both the prompt and displaced analyses, projected for both the Run~2 data set and the HL-LHC.
Shaded bands indicate variations in the limit corresponding to uncertainties for the prompt search and variations in the signal efficiency for the displaced search, described in Sections~\ref{sec:stats} and \ref{sec:analysisDisp}, respectively.
Because the \zdark\ is very light in this scenario and thus considerably boosted, the displaced analysis is dominant, with sensitivity of the prompt analysis being similar to existing constraints.

\begin{figure}[t!]
\centering
\includegraphics[width=0.75\textwidth]{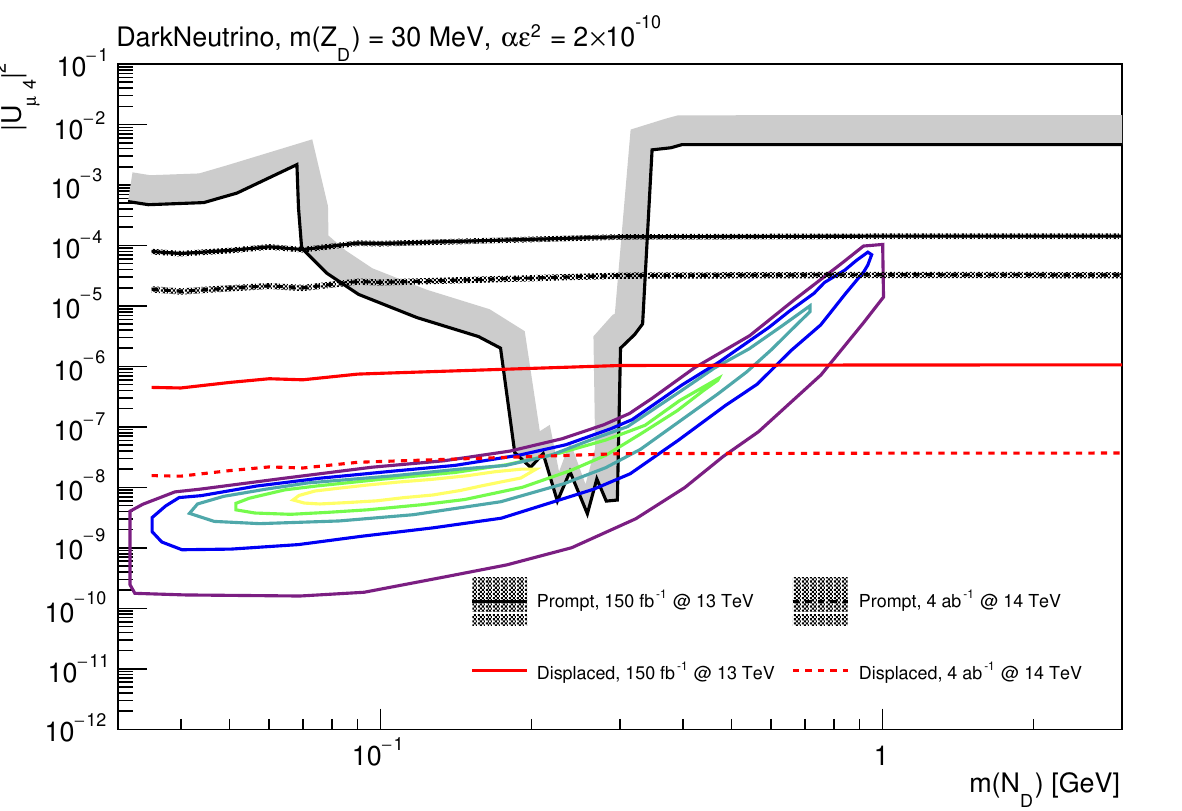}
\caption{The ranges of neutrino mixing parameters expected to be excluded by LHC experiments during Run~2 and the HL-LHC are shown versus \ndark\ mass for a fixed \zdark\ mass of 30 MeV. Upper limits on \UmFour\ are shown for both the prompt and displaced search strategies, with uncertainty bands described in the text.
Closed contours correspond to the parameters favored by Ref.~\cite{Bertuzzo:2018itn} at the 1, 2, 3, 4 and 5-sigma confidence level and shaded grey regions correspond to prior constraints, described in the text.}
\label{fig:limitNominal1}
\end{figure}

Figure~\ref{fig:limitNominal2} shows excluded regions of \zdark\ and \UmFour\ under the assumption of $m_{\ndark}=3m_{\zdark}$.
Expected limits are shown for both the prompt and displaced analysis strategies, separately for integrated luminosities and energies corresponding to Run~2 and the HL-LHC.
The proper branching fractions of the \zdark\ are considered for the electron and muon channels, which are also statistically combined.
For low \zdark\ masses, larger lifetimes and boost factors cause the displaced analysis to contribute the dominant sensitivity while at larger masses, the prompt analysis yields the strongest limit.

\begin{figure}[t!]
\centering
\includegraphics[width=0.75\textwidth]{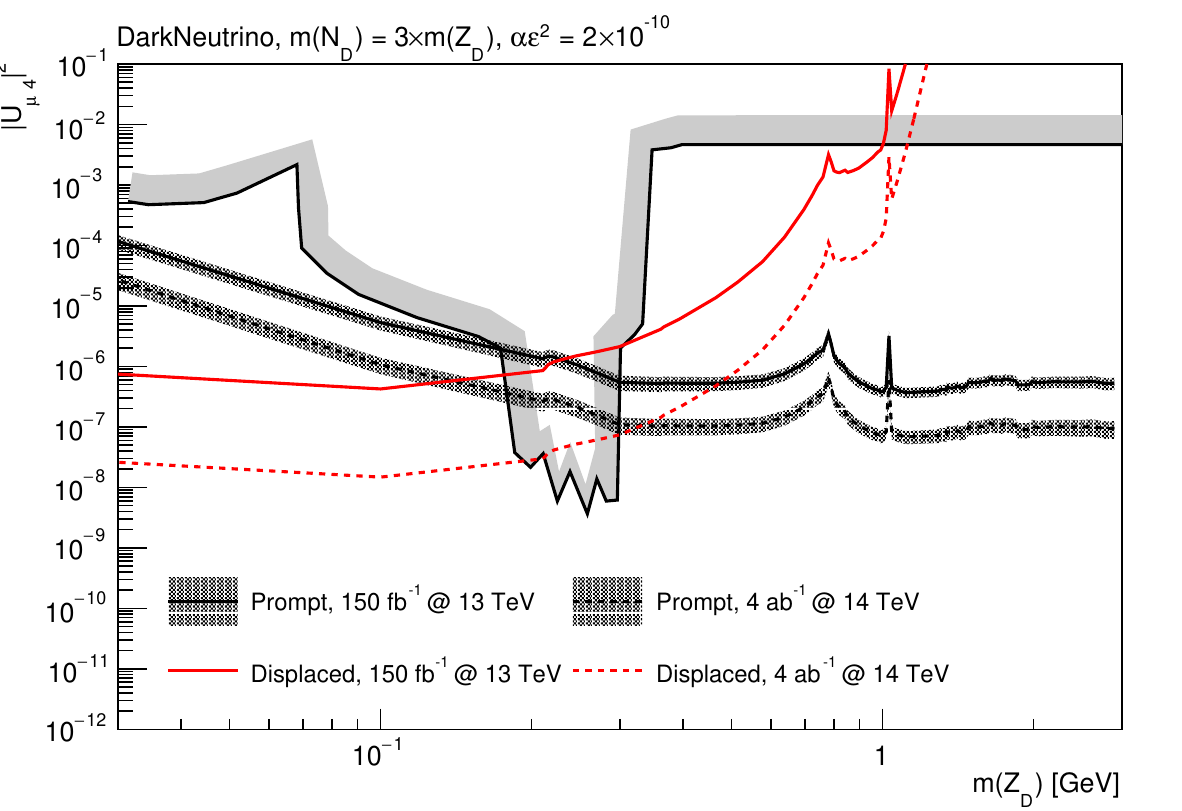}
\caption{Expected upper limits on the neutrino mixing parameter \UmFour\ are shown as a function of the \zdark\ mass under the assumption that $m_{\ndark} = 3 m_{\zdark}$. 
The results of the prompt and displaced analysis are separately presented, assuming both thte Run~2 and HL-LHC data sets.}
\label{fig:limitNominal2}
\end{figure}

Differences in the value of kinematic mixing parameter $\epsilon$ can give rise to different expected distributions of the displaced vertex displacement for a fixed \zdark\ mass.
The impact of these differences are shown in Figure~\ref{fig:limitEpsilons}, where the limits expected for the nominal value of $\alpha\epsilon^2=\times10^{-10}$ are compared to those for values of $\alpha\epsilon^2=\times10^{-9}$ and $\alpha\epsilon^2=\times10^{-11}$.
For the scenario with a 30~\,MeV \zdark\ mass, the displaced analysis continues to dominate in all cases, with the expected constraint varying based on the expected fraction of signal falling into the 1~\.mm to 10~\,cm search window.
For the scenario with  $m_{\ndark} = 3 m_{\zdark}$, an upward variation of $\epsilon$ generally will strengthen the limit from the prompt search and weaken the limit from the displaced one and vice-versa.
However, for very light \zdark\ the nominal choice of $\epsilon$ leads to a signal peaking outside of the acceptance considered for the displaced analysis, so consequently larger values of the parameter will improve the limit for both the prompt and displaced search.

\begin{figure}[t!]
\centering
\includegraphics[width=0.49\textwidth]{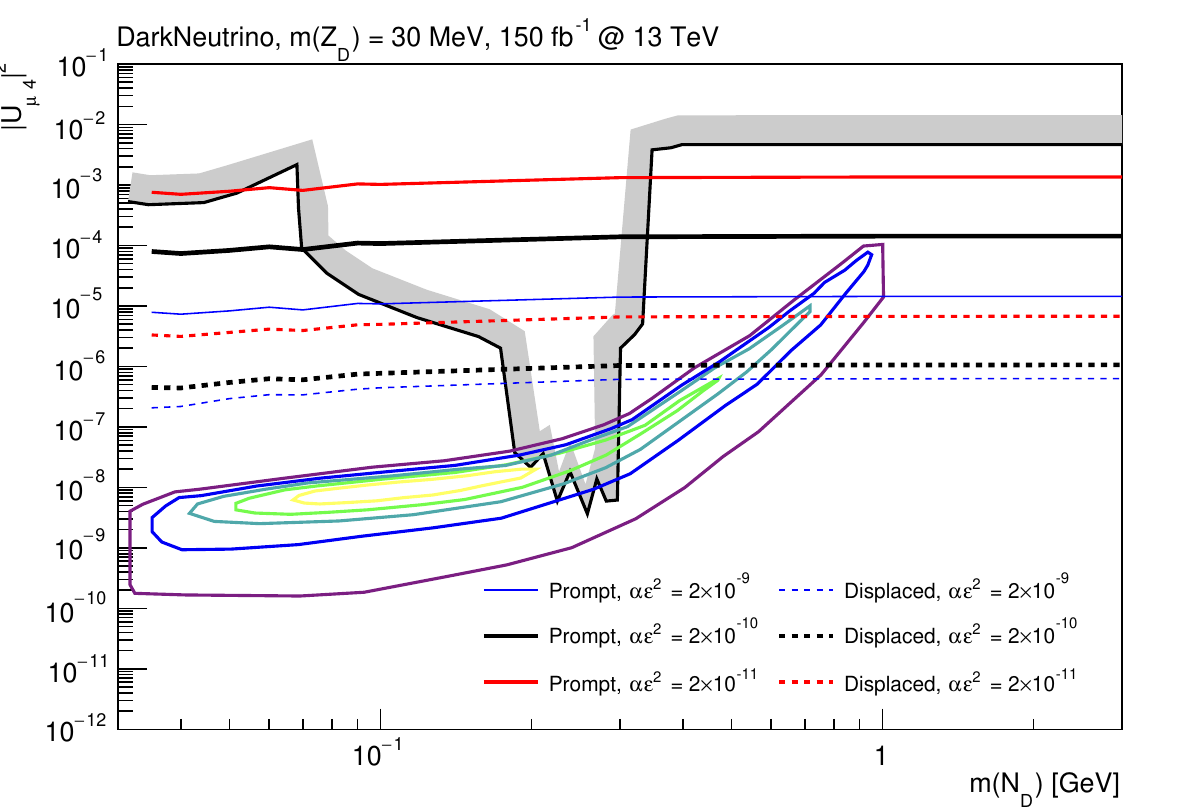}
\includegraphics[width=0.49\textwidth]{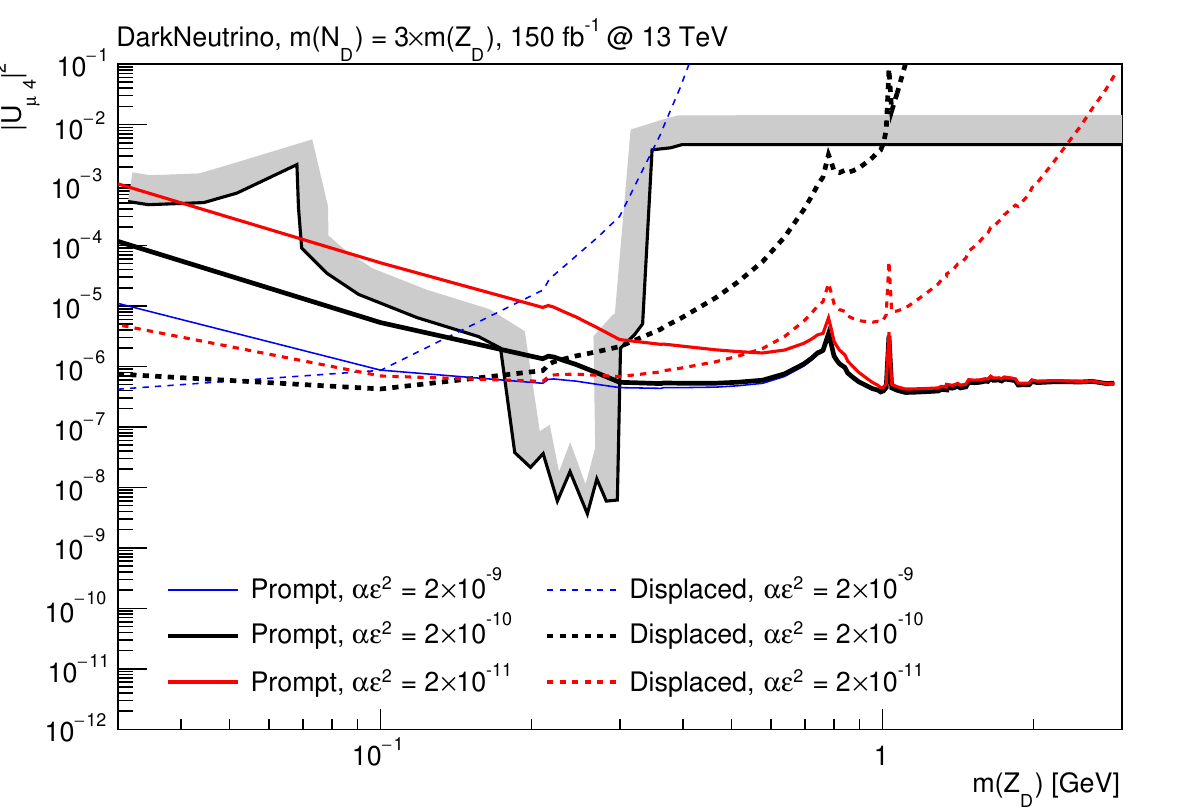}
\includegraphics[width=0.49\textwidth]{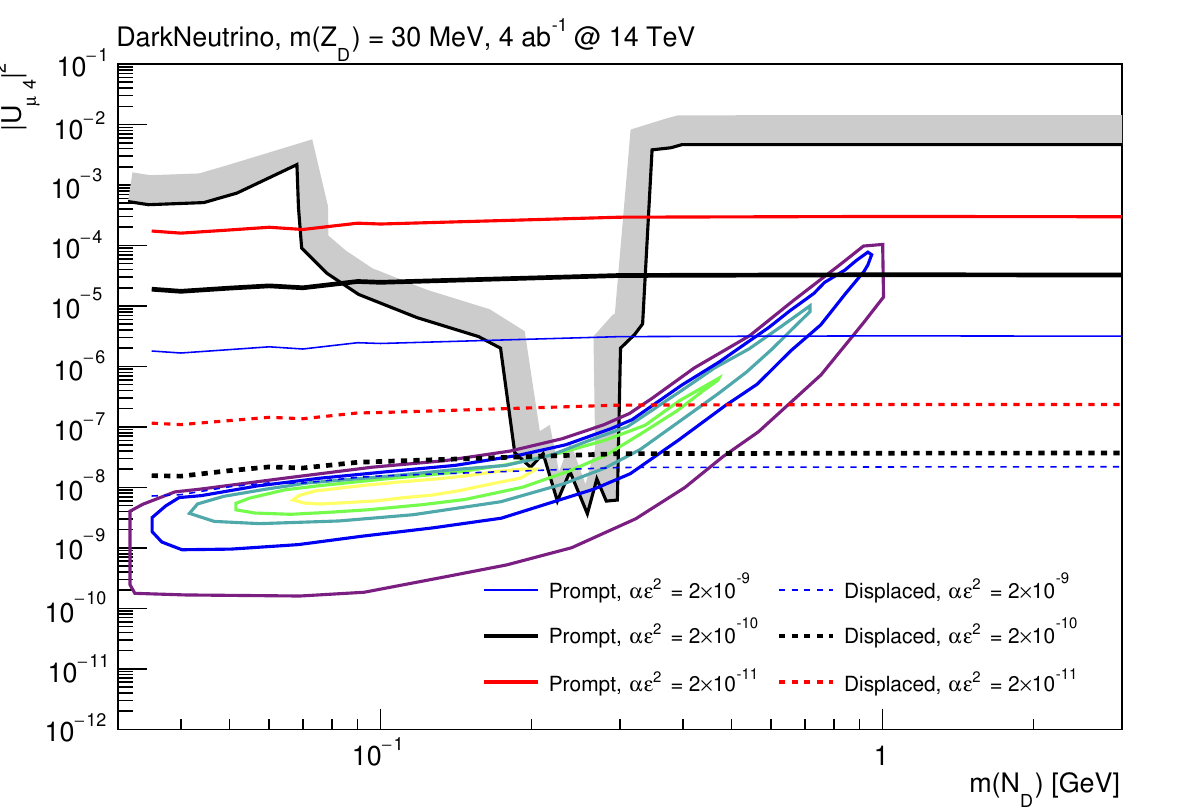}
\includegraphics[width=0.49\textwidth]{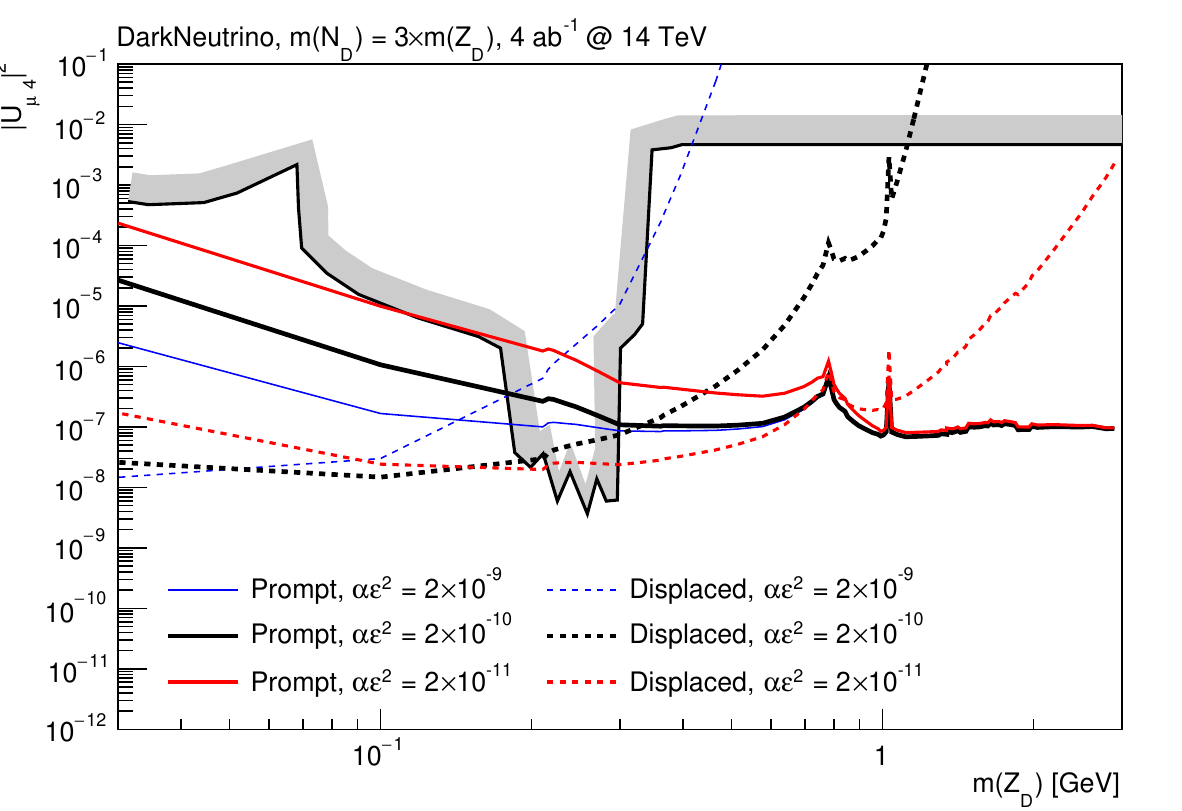}
\caption{Expected upper limits on the neutrino mixing parameter \UmFour\ are shown as a function of the \zdark\ mass for several assumptions on the kinetic mixing parameter.
Comparisons are made for the Run~2 (top row) and HL-LHC data sets (bottom row), and for the $m_{\zdark}=30$ MeV (left column) and $m_{\ndark} = 3 m_{\zdark}$ (right column) assumptions.
In each case, the results of both the prompt and displaced searches are shown, assuming values of $\alpha\epsilon^2=\times10^{-9}$, $\alpha\epsilon^2=\times10^{-10}$, and $\alpha\epsilon^2=\times10^{-11}$.}
\label{fig:limitEpsilons}
\end{figure}

The expected lepton \pt\ spectra depends on the specific signal mass parameters under consideration, but generally peaks at low values, motivating the use of minimal thresholds for analysis.
Figure~\ref{fig:limitLeptons} compares various possibilities along these lines.
In the prompt analysis, the nominal scenario of 3~\,GeV muons and 5~\,GeV electrons is augmented by expected limits that would result from lowering the electron \pt\ threshold to 1~\,GeV.
The displaced analysis nominally considers electrons and muons with \pt\ above 5~\,GeV, with a 1~\,GeV scenario also considered for electrons.
In all cases, limits for the combined electron and muon channels are computed by adding the expected significance for each channel in quadrature.

\begin{figure}[t!]
\centering
\includegraphics[width=0.49\textwidth]{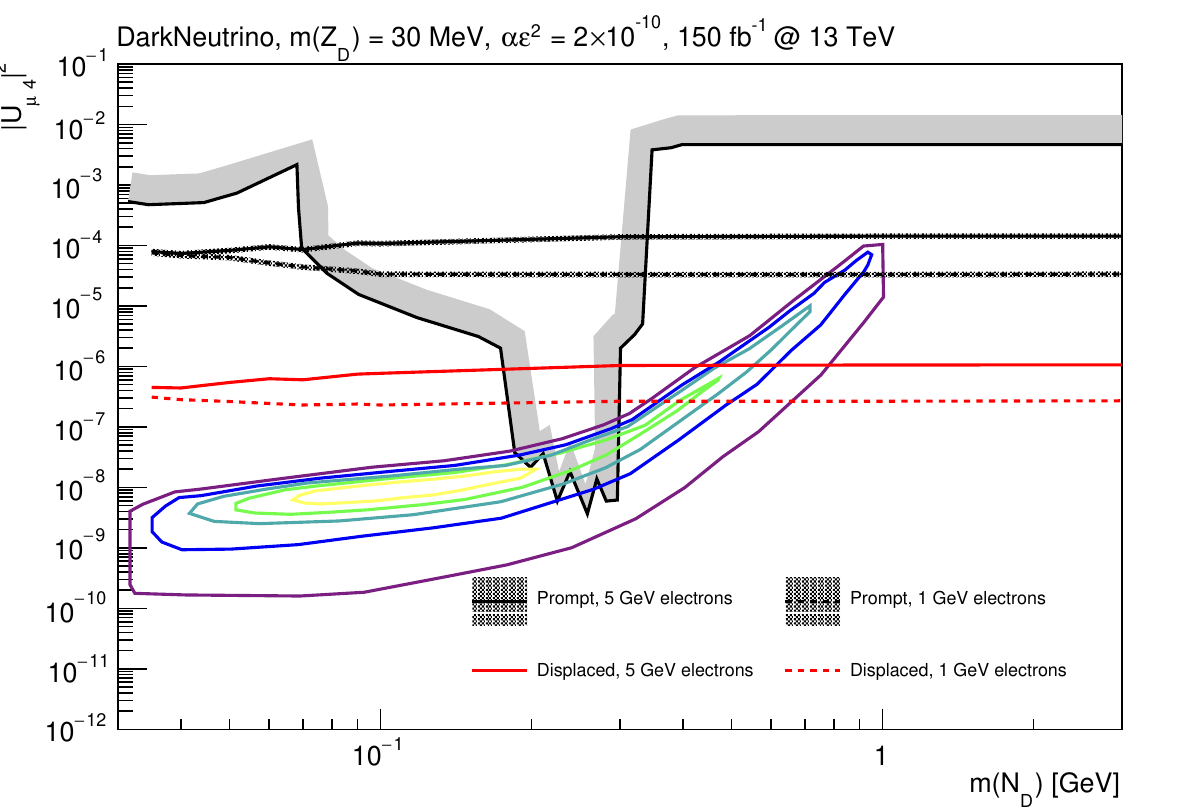}
\includegraphics[width=0.49\textwidth]{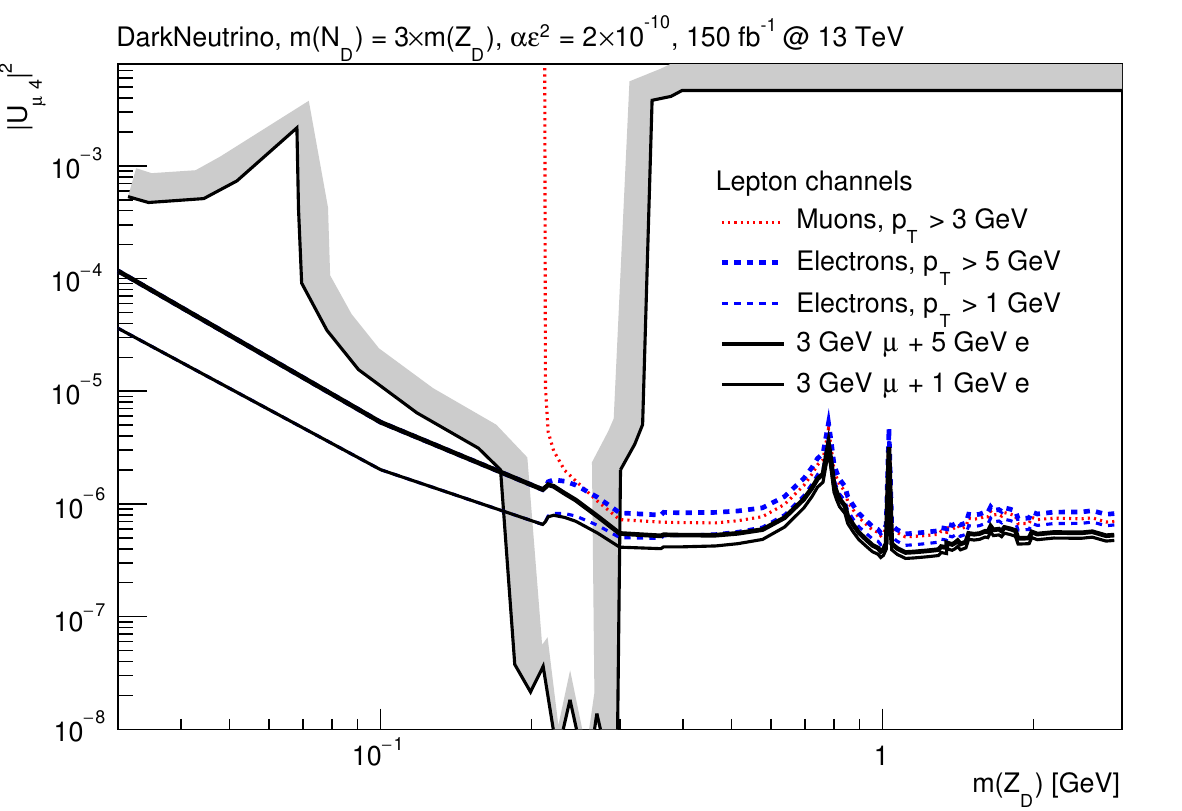}
\includegraphics[width=0.49\textwidth]{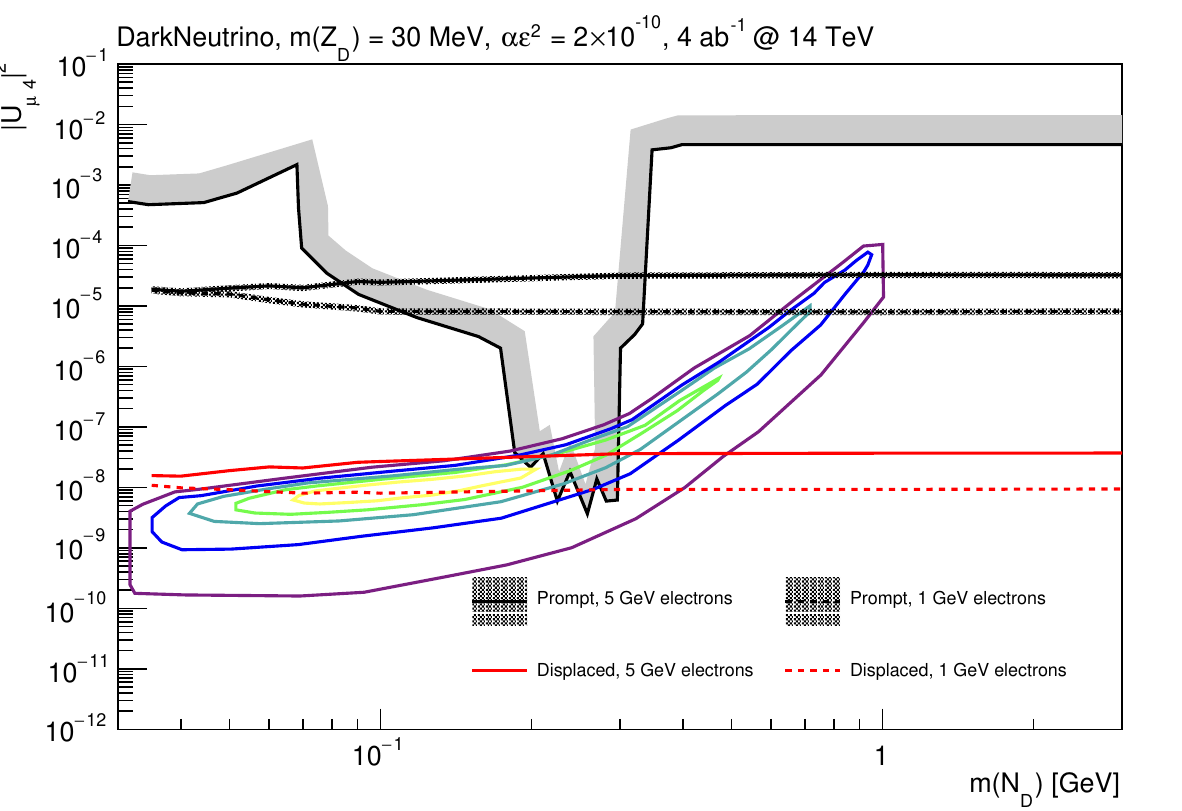}
\includegraphics[width=0.49\textwidth]{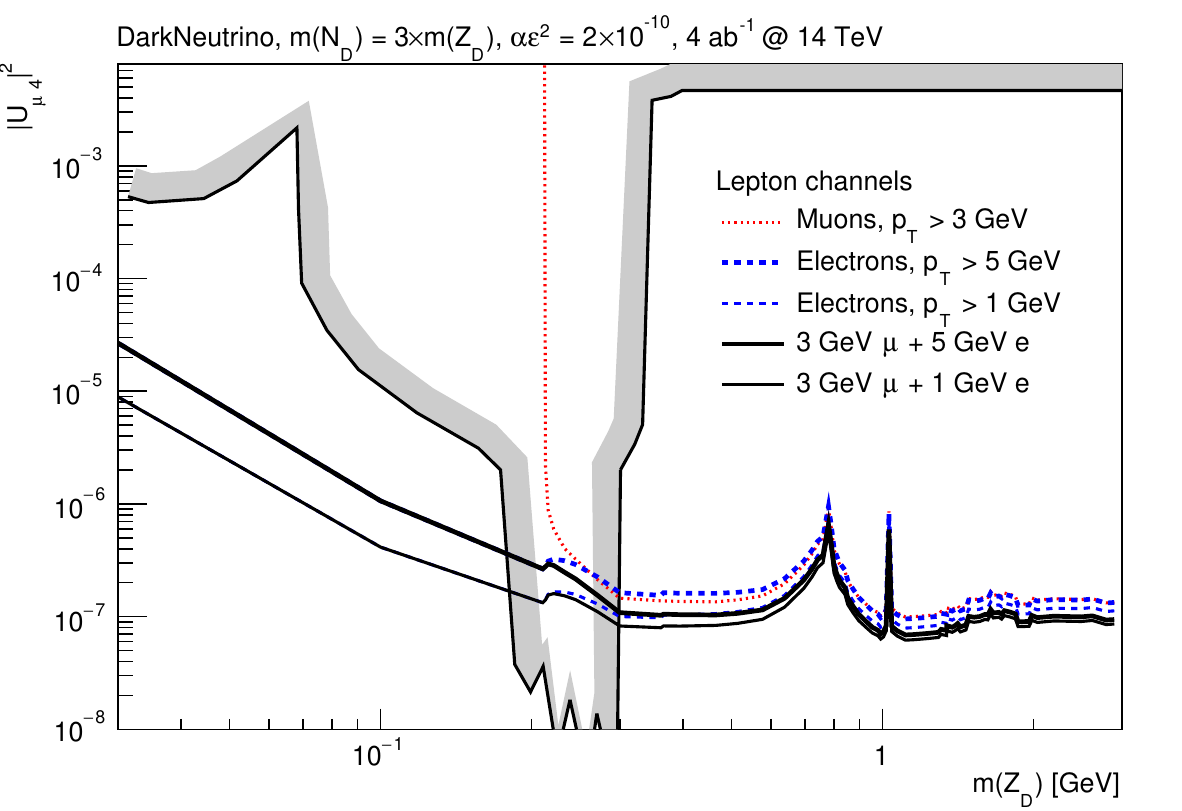}
\caption{Expected upper limits on the neutrino mixing parameter \UmFour\ are shown as a function of the \zdark\ mass for several assumptions about the classes of potentially reconstructable leptons.
Comparisons are made for the Run~2 (top row) and HL-LHC data sets (bottom row), and for the $m_{\zdark}=30$ MeV (left column) and $m_{\ndark} = 3 m_{\zdark}$ (right column) assumptions.}
\label{fig:limitLeptons}
\end{figure}

\section{Conclusions and Outlook}
\label{sec:outlook}

Despite many significant efforts to understand the source of the excess of low-energy events seen by the MiniBooNE, a satisfying solution remains elusive.
The dark neutrino model offers an exciting possibility that this anomaly could be our first hint of a rich new sector of particles, sterile under the SM gauge forces, but with its own gauge symmetry and scalar sector.
We have proposed a new method to probe this compelling scenario using high-energy neutrinos from the large samples of $W$ bosons collected by the ATLAS and CMS experiments at the LHC.
Search strategies considering both prompt and displaced lepton pairs should each be pursued and contribute complementary sensitivities that depend on the particular parameters of the signal model.
The coverage of these strategies is shown to be highly complementary with low-energy experiments, reaching mixing parameters $|\UmFour|^2$ from $10^{-6}$ to $10^{-8}$ across a broad range of dark neutrino masses.
Notably, very low \zdark\ masses, accessible only via the displaced decays to electron pairs, can be explored thanks to the large Lorentz boost factor afforded by the W boson production channel.
Unique sensitivity can be achieved with the $\sqrt{s}=13\Tev$ data that has already been collected, and the High-Luminosity LHC will provide an unprecedented sample of $W$ decays to probe dark neutrino production with exceeding-small portal couplings.

\vspace{0.5cm}
\noindent \textbf{Acknowledgments}: We warmly thank Anadi Canepa for making the initial connections that lead to this work. We also thank Stefan H{\"o}che and Frank Krauss for useful conversations. The authors' work is supported by Fermi Research Alliance, LLC under Contract No. DE-AC02-07CH11359 with the U.S. Department of Energy, Office of Science, Office of High Energy Physics. ARH is supported by the U.S. Department of Energy, Office of Science, Office of High Energy Physics under Award Number 89243023SSC000116.

\bibliography{bibliography}

\end{document}